\documentclass[12pt]{article}

\usepackage{graphicx,psfrag,epsf,color}
\usepackage{amsmath,amssymb,amsfonts}
\usepackage{array}
\usepackage{cite}
\usepackage{rotating}
\usepackage{slashed, cancel}
\newcounter{MBQ}

\newcommand{\be}{\begin{equation}}
\newcommand{\ee}{\end{equation}}
\newcommand{\bea}{\begin{eqnarray}}
\newcommand{\eea}{\end{eqnarray}}
\newcommand{\bi}{\begin{itemize}}
\newcommand{\ei}{\end{itemize}}
\newcommand{\ben}{\begin{enumerate}}
\newcommand{\een}{\end{enumerate}}
\newcommand{\bt}{\begin{tabular}}
\newcommand{\et}{\end{tabular}}

\newcommand{\nn}{\nonumber}

\newcommand{\nm}{n_-}
\newcommand{\np}{n_+}

\newcommand{\T}{{\bf T}}

\newcommand{\nnm}[1]{n_{#1-}}
\newcommand{\nnp}[1]{n_{#1+}}

\newcommand{\ci}{c}

\newcommand{\als}{\alpha_s}

\global\long\def\order#1{\mathcal{O}\left(#1\right)}
\newcommand{\Aus}{A_{\textrm{s}}} 

\begin{document}
\allowdisplaybreaks

\begin{titlepage}

\begin{flushright}
{\small
TUM-HEP-1116/17\\
December 12, 2017
}
\end{flushright}

\vskip1cm
\begin{center}
{\Large \bf\boldmath Anomalous dimension of subleading-power\\[0.2cm] 
${N}$-jet operators}
\end{center}

\vspace{0.5cm}
\begin{center}
{\sc Martin~Beneke, Mathias~Garny, Robert~Szafron, Jian~Wang} \\[6mm]
{\it Physik Department T31,\\
James-Franck-Stra\ss e~1, 
Technische Universit\"at M\"unchen,\\
D--85748 Garching, Germany\\
}
\end{center}

\vspace{0.6cm}
\begin{abstract}
\vskip0.2cm\noindent
We begin a systematic investigation of the anomalous dimension of 
subleading power $N$-jet operators in view of resummation of 
logarithmically enhanced terms in partonic cross sections beyond 
leading power. We provide an explicit result at the one-loop 
order for fermion-number two $N$-jet operators at the second order in 
the power expansion parameter of soft-collinear effective theory.
\end{abstract}
\end{titlepage}

\section{Introduction}
\label{sec:introduction}

The scattering amplitude of $N$ well-separated, energetic, massless 
particles is one of the key quantities in gauge theories. Understanding 
its structure is of fundamental importance, both for its own reason 
by revealing mathematical structure that is not at all evident from 
the underlying Lagrangian and its Feynman rules, and for the phenomenology 
of high-energy scattering in QCD.

Of particular interest are the soft and collinear divergences, which 
exhibit a high degree of universality. Some form of analytic calculation 
is usually required in order to efficiently cancel the divergences 
between virtual and real emission effects in infrared-safe scattering 
cross sections. The infrared divergences of the virtual $N$-parton 
scattering amplitude are governed by the soft-collinear anomalous 
dimension, which up to the two-loop order has the very simple structure 
\begin{equation}
\mathbf{\Gamma} = -\gamma_{\rm cusp}(\alpha_s) \sum_{i<j} 
\mathbf{T}_i\cdot\mathbf{T}_j \ln\left(\frac{-s_{ij}}{\mu^2}\right) 
+ \sum_{i}
\gamma_i(\alpha_s) 
\label{eq:LPanomalousdim}
\end{equation}
in colour-operator notation \cite{Catani:1998bh} and for all out-going 
momenta $p_i$ with $s_{ij}=2 p_i\cdot p_j+i0$, $i,j=1\ldots N$. The soft 
\cite{Almelid:2015jia} and collinear \cite{Moch:2005id}
contributions to $\mathbf{\Gamma}$ are 
known up to the three-loop order.\footnote{We refer to the above papers 
for a comprehensive list of references to relevant results at lower 
orders.} The above assumes that all scalar products $s_{ij}$ are 
parametrically of the same order as some hard scale $Q$. If the 
physical observable is sensitive to a smaller scale $M$ generated 
by soft or collinear radiation, the anomalous dimension is a central 
object in the systematic all-order resummation of large logarithms 
$\ln Q/M$ in the expansion in the coupling $\alpha_s$.

When this is the case the above anomalous dimension refers to the 
infrared singularities at leading order in the expansion in powers 
of $M/Q$ (``leading power''). Given the advances in the understanding 
of multi-loop corrections to the leading power anomalous dimension, 
it is also timely to ask about the next, subleading power term in 
the $M/Q$ expansion. It has been known for a long-time that single soft 
emission from an $N$-jet amplitude is described by a universal 
expression, the LBK amplitude, also at next-to-leading 
power\cite{Low:1958sn,Burnett:1967km}. This result extends the eikonal 
formula and has recently attracted new interest in connection with 
a possible relation to an asymptotic symmetry at null 
infinity \cite{Strominger:2013lka}. However, little is known about the 
structure of divergences of loops and the anomalous dimension at the 
subleading powers. The exponentiation of purely soft, ``next-to-eikonal'' 
effects has been discussed in Refs.~\cite{Laenen:2008gt,Laenen:2010uz}.
However, a major complication at next-to-leading power arises from 
the interplay of soft and collinear radiation 
as can be seen, for example, from the failure (or rather---depending on the 
point of view---generalization) of the LBK formula for jet processes 
beyond the tree approximation~\cite{DelDuca:1990gz,Larkoski:2014bxa}.

In this paper we begin with a systematic study of subleading power 
$N$-jet operators and their anomalous dimension with the ultimate goal 
of being able to sum logarithmically enhanced loop effects to all 
orders in perturbation theory. We base this study on soft-collinear 
effective theory 
(SCET)~\cite{Bauer:2000yr,Bauer:2001yt,Beneke:2002ph,Beneke:2002ni}, 
which offers the advantage that the power counting required to identify 
all next-to-leading power terms is already built into the Lagrangian. 
While we will not solve the resummation problem here and do not even 
discuss logarithms for a physical process, our approach demonstrates 
a clear path how this could be done in principle and systematically. 
The structure of the anomalous dimension matrix of subleading-power $N$-jet 
operators will become apparent and we provide the first complete 
result for the class of fermion-number $F=2$ operators to begin with.
Previous work on anomalous dimensions 
at subleading power in SCET focused on specific cases, the heavy-to-light 
current \cite{Hill:2004if, Beneke:2005gs} (related to 
$J^{B1}_{{\cal A}\chi}$ in the operator basis defined below) 
in the position-space SCET formalism, 
and on power-suppressed tree-level currents relevant to $e^+ e^-\to$~two 
jets in a different SCET framework~\cite{Freedman:2014uta,Goerke:2017lei}.

Several other works have recently addressed next-to-leading power (NLP)
effects from a more practical perspective. In 
Refs.~\cite{Bonocore:2014wua,Bonocore:2015esa,Bonocore:2016awd} the 
threshold limit of the partonic Drell-Yan process has been 
investigated and all NLP terms of the 
next-to-next-to-leading order (NNLO) cross section have been 
successfully reproduced in a diagrammatic expansion analysis. Also 
a ``radiative jet function'' has been identified, related 
to collinear effects, which appear near threshold first at NLP. 
For colourless final states the interference of the NLP LBK 
amplitude with the tree process allows one to compute the  
NLP terms at NLO in the loop expansion \cite{DelDuca:2017twk}. 
Another recent development \cite{Moult:2016fqy,Boughezal:2016zws} 
concerns the analytic computation of the leading NLP logarithm at NNLO 
in the separation parameter of the $N$-jettiness subtraction 
method \cite{Boughezal:2015dva,Gaunt:2015pea}, making the cancellation 
of the dependence on the separation parameter in the full simulation 
of the process more efficient. All of these applications have in 
common that they refer at present to logarithms at fixed order 
in perturbation theory up to NNLO and to processes with only two 
collinear directions. The general approach outlined in the present paper, 
once developed, should allow the computation of further logarithms 
in these applications, and in particular their resummation to all 
orders. We finally take note that along a somewhat different direction 
a formula for fermion-mass suppressed double logarithms in the high-energy 
limit of certain fermion-scattering form factors has been 
derived~\cite{Penin:2014msa,Liu:2017vkm}.

\section{\boldmath Subleading $N$-jet operator basis}
\label{sec:basis}

It was noted in Refs.~\cite{Becher:2009cu,Becher:2009qa} that the infrared 
anomalous dimension (\ref{eq:LPanomalousdim}) must correspond to the 
ultraviolet divergences of soft and collinear loops in SCET, if SCET is 
to be the correct effective field theory for jet processes. This observation 
also applies to subleading powers. The following analysis is based 
on the position-space field representation of 
SCET \cite{Beneke:2002ph,Beneke:2002ni}. The physical processes which 
are covered by this analysis are those for which the virtuality of 
collinear modes in any of the $N$ jet directions is of the same order, 
and parametrically larger than the one of the soft mode. The power-counting 
parameter $\lambda$ is set by the transverse momentum $p_{\perp i} 
\sim Q\lambda$ of collinear momenta with virtuality 
${\cal O}(\lambda^2)$.\footnote{$Q$ denotes a generic large energy/hard 
scale, which we set to 1 in the following.} 
The components of soft momentum are all ${\cal O}(\lambda^2)$ and 
consequently soft virtuality scales as $\lambda^4$. Below the term 
``NLP'' refers to ${\cal O}(\lambda)$ and ${\cal O}(\lambda^2)$, since 
the first non-vanishing power correction to most physical processes of 
interest is ${\cal O}(\lambda^2)$.

Under these assumptions (often referred to as SCET$_{\rm I}$) the 
SCET Lagrangian including all subleading power interactions to 
${\cal O}(\lambda^2)$ was already given in Ref.~\cite{Beneke:2002ni}. For 
$N$ widely separated collinear directions, the Lagrangian 
\begin{equation}
{\cal L}_{\rm SCET} = \sum_{i=1}^N {\cal L}_i(\psi_i,\psi_s) 
+{\cal L}_s(\psi_s)
\label{eq:SCETI}
\end{equation}
is the sum of $N$ copies of collinear Lagrangians with $N$ pairs of separate 
light-like reference vectors $n_{i \pm }$, $i=1,\dots,N$ satisfying 
$\nnm{i}\cdot\nnm{j} = {\cal O}(1)$. The collinear fields $\psi_i$ all 
interact with the same soft field $\psi_s$ but not among each other. 
The SCET Lagrangian is invariant under $N$ separate collinear 
gauge transformations and a soft gauge transformation, 
see Ref.~\cite{Beneke:2002ni}.

We therefore proceed to the construction of a complete basis of subleading 
$N$-jet operators in SCET. The general structure 
\be
J = \int dt\; C(\{t_{i_k}\}) \,J_s(0) 
\prod_{i=1}^N J_i(t_{i_1},t_{i_2},\dots)
\label{eq:Njetop}
\ee
can be described by products of operators $J_i$ associated to collinear 
directions $\nnp{i}$,  
each of which is itself composed of a product of $n_i$ gauge-invariant 
collinear ``building blocks'' $\psi_{i_k}$ \cite{Beneke:2004in},
\be
J_i(t_{i_1},t_{i_2},\dots) = \prod_{k=1}^{n_i} \psi_{i_k}(t_{i_k}\nnp{i})\,,
\ee
and a soft operator $J_s$. In general, each of the collinear building blocks 
is integrated over the corresponding collinear direction in position space,
where $C(\{t_{i_k}\})$ is a Wilson coefficient, and $dt=\prod_{ik} dt_{i_k}$.
Apart from the displacement along each of the collinear directions, the 
operators are evaluated at position $X=0$, corresponding to the location of 
the hard interaction.

The guiding principle for constructing building blocks is the requirement of collinear and soft gauge covariance. Because each collinear sector transforms under its own collinear gauge transformation, each collinear building block must be a collinear gauge singlet. However, the soft field may interact with different collinear sectors so we only need to assume that collinear building blocks transform covariantly under the soft gauge transformation. 
Note that, in general, the collinear building blocks may also contain multipole expanded soft fields.
For a collinear block the transformation properties under collinear and soft gauge transformation may be summarized as follows  
\be
J_i(x) \xrightarrow{\textrm{coll.}} J_i(x),\qquad
J_i(x) \xrightarrow{\textrm{soft}} U_s(x_{i-}) J_i(x)\,, 
\ee
where $x_{i-}^\mu = (\nnp{i}x)\,\nnm{i}^\mu/2$ and $U_s$ refers to the 
(not necessarily irreducible) colour representation of $J_i$. For the 
matrix adjoint representation we would have $J_i(x) 
\xrightarrow{\text{soft}} U_s(x_{i -}) J_i(x) U^\dagger_s(x_{i-})$ with 
$U_s$ in the fundamental representation. 

The elementary collinear-gauge-invariant collinear building blocks are 
given by 
\be
\psi_{i}(t_{i}\nnp{i}) \in 
\left\{ \begin{array}{ll} 
\chi_i(t_{i}\nnp{i}) \equiv W_{i}^\dag\xi_{i} & 
\hspace*{0.5cm} \mbox{collinear quark}\\[0.2cm] 
{\cal A}_{\perp i}^\mu(t_{i}\nnp{i})\equiv 
W_{i}^\dag [iD_{\perp i}^\mu W_{i}] & 
\hspace*{0.5cm} \mbox{collinear gluon} 
\end{array}\right.
\label{eq:elementaryblock}
\ee
for the collinear quark and gluon field in $i$-th direction, respectively. 
$W_i$ is the path-ordered exponential of $\nnp{i} A_{i}$ 
(``$i$-collinear Wilson line'') and the covariant derivative includes 
only the collinear gluon field. Both, the quark and gluon building blocks 
scale as ${\cal O}(\lambda)$~\cite{Beneke:2002ph}. 
Objects containing $i\nnp{i} D_i$ 
or $i \nnp{i}\partial$ are redundant. The first can be reduced to the second 
with the help of $i\nnp{i}D_iW_i = W_i i\nnp{i}\partial$ and 
$W_i^\dag i\nnp{i} D_i = i\nnp{i}\partial W_i^\dag$.\footnote{
Covariant derivatives acting on Wilson lines are understood as operators 
acting on functions to the right. When the derivative should be understood 
to operate only on the Wilson line, we add a square bracket as in 
Eq.~(\ref{eq:elementaryblock}) for clarity. In all other cases the 
derivative is meant to act only on whatever is written explicitly to 
the right or within brackets.} 
The ordinary derivatives 
can be removed using $i\nnp{i}\partial\psi_{i_k}(t_{i_k}\nnp{i}) 
=i d\psi_{i_k}(t_{i_k}\nnp{i})/dt_{i_k}$ followed by an 
integration by parts in the $t_{i_k}$-integral in Eq.~(\ref{eq:Njetop}).

At leading power, only a single building block contributes for each 
direction, i.e.~$n_i=1$ for all $i=1,\dots,N$,
and the elementary building blocks are given by 
\be
J_i^{A0}(t_{i}) = \psi_{i}(t_{i}\nnp{i})\,.
\ee
The superscript in $J_i^{A0}$ indicates the leading-power contribution, and 
the reason for this nomenclature will become clear in a moment.
We are interested in $N$-jet operators that are suppressed by one or two 
powers of $\lambda$ relative to the leading power. This suppression can 
arise in three ways: 
\begin{itemize}
\item[(i)] via higher-derivative operators, i.e. acting with either 
$i\partial_{\perp i}^\mu\sim{\cal O}(\lambda)$ or 
$i\nnm{i}D_s\equiv i\nnm{i}\partial+g_s\nnm{i}A_s(x_{i-})
\sim{\cal O}(\lambda^2)$ on the elementary building blocks 
$\psi_{i_k}$. Here it is important to note that since the elementary building 
blocks transform under the soft gauge transformation with $U_s(x_{i-})$, 
the covariant soft derivative is the ordinary derivative for the transverse 
direction and $i\nnm{i}D_s$ for the $\nnm{i}$ projection. In other words, 
the soft covariant derivative {\em on collinear building blocks} 
is $iD_s^\mu(x) \equiv i\partial^\mu + 
g_s\nnm{i}\Aus(x_{i-})\frac{\nnp{i}^\mu}{2}$ due to the multipole 
expansion of the soft fields, which  guarantees a homogeneous scaling 
in $\lambda$;
\item[(ii)] by adding more building blocks in a given direction, i.e. 
$n_i>1$, since $ \chi_i\sim {\cal O}(\lambda)$ and 
${\cal A}_{\perp i}^\mu \sim {\cal O}(\lambda)$,
\item[(iii)] via new elementary building blocks that appear at 
subleading power, including purely soft building blocks in $J_s$.
\end{itemize}
In the following, we label operators that consist of a single building block
by $J_i^{An}$, where $n=1,2$ indicates the relative power suppression due to 
additional derivatives. Using the equation of motion derived from the leading 
power collinear Lagrangian, it is possible to eliminate operators with 
$i\nnm{i}D_s$ derivatives (see below and App.\,\ref{sec:redundant}), such that 
the operator basis consists of
\bea
J_i^{A1}(t_{i}) &=& i\partial_{\perp i}^\nu J_i^{A0} 
\qquad\quad {\cal O}(\lambda)\,,\\
J_i^{A2}(t_{i}) &=& 
i\partial_{\perp i}^\nu\,i\partial_{\perp i}^\rho J_i^{A0} 
\quad {\cal O}(\lambda^2)\,. \label{eq:A2}
\eea
Covariant derivative operators such as 
$(W_{i}^\dag iD_{\perp i}^\mu\xi_{i})(t_i\nnp{i})$ and 
$(W_{i}^\dag iD_{\perp i}^\mu  iD_{\perp i}^\nu W_{i})(t_i\nnp{i})$ are 
special cases of $J_i^{A1}(t_{i})$ and the 
$J_i^{B1}(t_{i_1},t_{i_2})$ defined in the following 
with $t_{i_1}=t_{i_2}$. Hence all derivative basis operators are constructed 
from ordinary transverse derivatives acting on gauge-invariant collinear 
building blocks.

Operators with two collinear building blocks in the same direction $i$ 
are suppressed at least by one power of $\lambda$ with respect to the leading 
power, and we label them by $J_i^{Bn}$. At ${\cal O}(\lambda)$,
\be
J_i^{B1}(t_{i_1},t_{i_2}) = 
\psi_{i_1}(t_{i_1}\nnp{i})\psi_{i_2}(t_{i_2}\nnp{i})
\in \left\{ \begin{array}{ll}
{\cal A}_{\perp i}^\mu(t_{i_1}\nnp{i}) \chi_i(t_{i_2}\nnp{i}) \\[0.03cm]
\chi_i(t_{i_1}\nnp{i})\chi_i(t_{i_2}\nnp{i}) \\[0.03cm]
{\cal A}_{\perp i}^\mu(t_{i_1}\nnp{i}) 
{\cal A}_{\perp i}^\nu(t_{i_2}\nnp{i}) \\[0.03cm] 
\chi_i(t_{i_1}\nnp{i}) \bar\chi_i(t_{i_2}\nnp{i}) \;.
\end{array}\right.
\ee
The first operator has fermion number one, the second two, and the last two 
have fermion number zero. We do not list explicitly the conjugate operators 
with negative fermion number.

At ${\cal O}(\lambda^2)$, the operators $J_i^{B2}$ are obtained by acting 
with a $\partial_{\perp i}^\mu$ derivative on $J_i^{B1}$.
We will use a basis where the derivative acts either on the second building 
block, or on both,
\be\label{eq:B2}
J_i^{B2}(t_{i_1},t_{i_2})  \in \left\{\begin{array}{l} 
\psi_{i_1}(t_{i_1}\nnp{i})i\partial_{\perp i}^\mu\psi_{i_2}(t_{i_2}\nnp{i})
\\[0.1cm]
i\partial_{\perp i}^\mu\big[\psi_{i_1}(t_{i_1}\nnp{i})
\psi_{i_2}(t_{i_2}\nnp{i})\big]\,,
\end{array}\right.
\ee 
where $\psi_{i_1}\psi_{i_2}$ can be any combination from $J_i^{B1}$.
Finally, at ${\cal O}(\lambda^2)$ it is possible to have operators composed 
of three elementary building blocks in a single
direction, which we collectively call $J_i^{C2}$,
\be\label{eq:C2}
J_i^{C2}(t_{i_1},t_{i_2},t_{i_3}) = \psi_{i_1}(t_{i_1}\nnp{i})
\psi_{i_2}(t_{i_2}\nnp{i})\psi_{i_3}(t_{i_3}\nnp{i})\;.
\ee 
This exhausts the options (i), (ii) from above at ${\cal O}(\lambda^2)$. 

An example for a new building block that scales as order $\lambda^2$ 
and hence could be used to construct ${\cal O}(\lambda)$ suppressed 
operators is 
\be\nnm{i}\mathcal{A}_{i} \equiv 
W_{i}^\dag i\nnm{i} D_{i} W_{i} -i \nnm{i} D_s
= W_{i}^\dag [i\nnm{i} D_{i} W_{i}] - g_s\nnm{i}A_s
\stackrel{\rm cLCG}{=} g_s\nnm{i}A_i\,,
\ee where soft gauge covariance requires that 
$i\nnm{i} D_{i}$ includes the collinear gluon and the multi\-pole-expanded 
soft gluon field. The subtraction term $-i \nnm{i} D_s$ in the second 
expression, which is also multipole expanded, is required to obtain a 
field rather than a differential operator, as is clear from the third 
expression, in which $i\nnm{i} D_{i}$ acts only within the square 
bracket.\footnote{Note that the collinear Wilson line transforms as 
$W_i\to U_c(x)W_i$ under collinear gauge transformations and 
$W_i\to U_s(x_-)W_iU_s(x_-)^\dag$ under soft gauge
transformations \cite{Beneke:2002ni}.} The last expression shows that 
in collinear light-cone gauge $\nnp{i}A_c = 0$ the new building block 
corresponds to the small component of the collinear gauge field. 
However, using the collinear-field equation of motion, we show in 
App.~\ref{sec:redundant} that $\nnm{i}\mathcal{A}_{i}$ can be expressed in 
terms of the elementary building blocks with only $\partial_{\perp i}$ 
derivatives, hence $\nnm{i}\mathcal{A}_{i}$ can be removed from the 
basis building blocks. As noted above for the transverse derivatives 
other possible placements of $i \nnm{i} D_i$ can always be reduced 
to (products of) existing objects. For example
\be W_i^\dagger i \nnm{i} D_i \xi_i = 
i\nnm{i}D_s\chi_{i}+\nnm{i}\mathcal{A}_{i}\,\chi_{i}\,,
\ee
\be
W_i^\dag ( i D_{\perp i}^\mu i \nnm{i} D_i W_i -  
i D_{\perp i}^\mu W_i i \nnm{i} D_s) = 
{\cal A}_{\perp i}^\mu\,\nnm{i}\mathcal{A}_{i}\,.
\ee

As already mentioned we show in App.~\ref{sec:redundant} that the 
$i \nnm{i} D_s$ soft covariant derivative, which operates on the elementary 
collinear building blocks in the form
\be
i \nnm{i} D_s \chi_i, \qquad [i \nnm{i} D_s, {\cal A}_{\perp i}^\mu], 
\ee
can be eliminated by equation-of-motion operator identities in terms of 
the A2, B2 and C2 structures defined in Eqs.~(\ref{eq:A2}),~(\ref{eq:B2}) 
and (\ref{eq:C2}).  This implies that $i \nnm{i} D_s$ can be eliminated 
from any collinear operator as 
\be
i \nnm{i} D_s(0) J_i(t_{i_1},t_{i_2},\dots) = 
\sum_{k=1}^{n_i}\psi_{i_1}(t_{i_1}\nnp{i})\ldots
[i \nnm{i} D_s(0) \psi_{i_k}(t_{i_k}\nnp{i})]
\ldots 
\psi_{i_{n_i}}(t_{i_{n_i}}\nnp{i})\,,\;
\ee
where the covariant derivative is understood in the colour representation 
of the object it operates on. Together with the above this implies 
that up to $\order{\lambda^2}$ we can use a basis 
of collinear building blocks that does not involve soft fields through 
covariant derivatives. It is constructed entirely from ordinary transverse 
derivatives and the elementary building block for the quark fields 
and the transverse gluon field.

In addition to the collinear building blocks, the $N$-jet operator may also contain a pure soft building block $J_s$. The soft fields do not transform under the collinear gauge transformation, such that $J_s$ is trivially a singlet under collinear gauge transformations. In the pure soft sector there is no need to perform the SCET multipole expansion of the soft fields and therefore the soft gauge transformation $U_s(x)$ in this case depends on $x$ rather than on $x_-$. The soft transformation of $J_s$ is 
\be
J_s(x) \xrightarrow{\textrm{coll.}} J_s(x),
\qquad J_s(x) \xrightarrow{\textrm{soft}} U_s(x) J_s(x)\,,
\ee
with $U_s$ taken in the appropriate representation. In the adjoint 
matrix representation we have $J_s(x) \xrightarrow{\text{soft}} 
U_s(x) J_s(x) U^\dagger_s(x)$ with $U_s(x)$ in the fundamental. 
The covariant pure soft building blocks start at $\order{\lambda^3}$, 
for example
\bea
 q(x) \sim \lambda^3,\qquad 
F_s^{\mu\nu} \sim \lambda^4, 
\qquad iD_s^\mu q(x) \sim \lambda^5\,, 
\eea
where {\em on soft building blocks} 
$iD_s^\mu(x)=i\partial^\mu+g_s A_s^\mu(x)$ and the soft field strength tensor 
is defined as $i g_s F_s^{\mu \nu}= [iD_s^\mu,iD_s^\nu] $. We can therefore 
drop $J_s(0)$ in Eq.~(\ref{eq:Njetop}) at $\mathcal{O}(\lambda^2)$. 
Therefore, soft fields enter neither via the soft nor via the collinear 
building blocks for our basis choice, up to ${\cal O}(\lambda^2)$.
This implies that the emission of a 
soft gluon from the hard process, which generates the $N$-jet operator, 
is entirely accounted for by Lagrangian interactions.

The case of $N$-jet operators differs from that of heavy-to-light currents, 
which consist of one collinear direction and a soft heavy-quark field, 
whose decay is the source of large energy for the collinear final state.
The basis of subleading SCET operators listed in 
Ref.~\cite{Beneke:2004in} does contain soft covariant derivatives 
at ${\cal O}(\lambda^2)$ due to the presence of the soft heavy-quark 
building block at leading power. The absence of soft 
building blocks in $N$-jet operators at ${\cal O}(\lambda^2)$ is also 
an important difference and simplification of the position-space 
vs. the label-field SCET formalism \cite{Bauer:2000yr,Bauer:2001yt}, where 
soft fields must be included in the basis operators at 
${\cal O}(\lambda^2)$~\cite{Larkoski:2014bxa,Feige:2017zci}. 
The difference arises from a different split into collinear and soft, 
since in the 
label formalism only the large and transverse component of collinear 
momentum are treated as labels, while the residual spatial dependence 
of all fields, collinear and soft,  is soft. The difference in the 
operator basis due to this is compensated by a corresponding difference 
in the soft-collinear interactions in the Lagrangian in the 
two formulations of SCET.

It is useful to consider Fourier transformation with respect to the 
positions $t_{i_k}$ in the collinear direction, 
\bea
J_i^{An}(P_i) &\equiv& P_i\int dt_{i}\, e^{-it_{i}P_i}\, J^{An}(t_{i})\,,
\nn \\
J_i^{Bn}(P_i,x_i) &\equiv& P_i^2\int dt_{i_1}dt_{i_2}\, 
e^{-i(t_{i_1}x_i+t_{i_2}\bar x_i)P_i}\, J^{Bn}(t_{i_1},t_{i_2})\,,
\nn \\
J_i^{Cn}(P_i,x_{i_1},x_{i_2}) &\equiv& P_i^3\int dt_{i_1}dt_{i_2}dt_{i_3}\, 
e^{-i(t_{i_1}x_{i_1}+t_{i_2}x_{i_2}+t_{i_3}x_{i_3})P_i}\, 
J^{Cn}(t_{i_1},t_{i_2},t_{i_3})
\label{eq:Fourier}
\eea
for operators with one, two and three building blocks, respectively, 
where $\bar x_i=1-x_i$, $x_{i_3}=1-x_{i_1}-x_{i_2}$ and $P_i$ is the total 
(outgoing) collinear momentum in direction $i$. Here  
we adopt the convention that $\nnp{i}p_{i_k} = x_{i_k}\nnp{i}P_i >0$ 
for an outgoing momentum in direction $i$, such that from 
Eq.~\eqref{eq:Fourier} also $P_i>0$ and $x_{i_k} \in [0,1]$ for all 
momenta outgoing, which we shall assume in the following.\footnote{
Equivalently, one could assume $\nnp{i}p>0$ for ingoing momenta. It is 
possible to translate between both cases by flipping the signs
$\nnp{i}\to -\nnp{i}$ and $\nnm{i}\to -\nnm{i}$ of all directions.
This sign change can be compensated by substituting $t_{i_k}\to -t_{i_k}$, 
such that the form of the building blocks in position space is unchanged. 
The only difference is then the sign in the exponents in 
Eq.~\eqref{eq:Fourier}, such that in collinear momentum space $P_i>0$ 
for ingoing momenta in that case. We do not consider here the situation 
where some momenta are ingoing and others are outgoing.}
In general, the basis of $N$-jet operators can then be written in the form
\be
  J(\{P_i\},\{x_{i_k}\}) = \prod_{i=1}^N J_i(P_i,x_{i_1},x_{i_2},\dots)
\ee
where $x_{i_k}$ are momentum fractions of the collinear momentum  in 
direction $i$, carried by the $k$-th building block. The operators are given 
by $J_i\in\{J_i^{An},J_i^{Bn},J_i^{Cn}\}$, depending on the number of 
collinear building blocks and the order in $\lambda$. For each direction $i$ 
one of the $x_{i_k}$  can be eliminated using the constraint 
$\sum_k x_{i_k}=1$, in accordance with the previous definitions. For brevity, 
we will omit the arguments $P_i$ indicating the total collinear momentum in 
direction $i$ if there is no danger of confusion, because it is conserved
in all processes we consider.

The total power suppression of the $N$-jet operator is then obtained from 
adding up the suppression factors in $\lambda$ from each direction.
For example, at ${\cal O}(\lambda^2)$, it is possible to either have a 
$J_i^{X2}$ operator (with $X=A,B,C$) in one direction and $J_i^{A0}$ 
operators in the remaining $N-1$ directions, or two operators 
$J_i^{X1}J_j^{Y1}$, with $X,Y=A,B$, and $J_i^{A0}$ operators in the remaining
$N-2$ directions. 

The infrared divergences of $N$-jet processes at NLP follow from the 
ultraviolet divergences of the matrix elements of the above operators 
computed with the SCET Lagrangian including NLP interactions. For the 
derivation of the anomalous dimension and renormalization group equation 
it is convenient to adopt the interaction picture and treat 
the subleading SCET Lagrangian as an interaction, such that all 
operator matrix elements are understood to be evaluated with the 
leading-power SCET Lagrangian. The basis of subleading power $N$-jet 
operators at a given order in $\lambda$ then includes further 
``non-local'' operators from the time-ordered products of the current 
operators $J$ at lower order in $\lambda$ with the subleading terms in 
the SCET Lagrangian. The ``local'' (in reality, light-cone) currents do 
not mix into the non-local time-ordered product operators, but the latter 
can, in principle, mix into the former. The non-local operators  
mix into themselves but the corresponding matrix of renormalization factors 
is given by the one for the local currents of lower order in $\lambda$ 
contained in the time-ordered product. The absence of further 
renormalization from the subleading soft-collinear interactions in 
the time-ordered product follows 
from the non-renormalization of the SCET Lagrangian to all orders in the 
strong coupling constant at any order in $\lambda$ \cite{Beneke:2002ph}.

At  ${\cal O}(\lambda)$ the time-ordered product operators are of the 
form 
\be
\label{eq:T1}
J^{T1}_i(t_i) = i \int d^4 x \,
T\left\{J^{A0}_i(t_i), {\mathcal L}_i^{(1)} (x)\right\}\,,
\ee
where ${\mathcal L}_i^{(1)} = {\mathcal L}_\xi^{(1)}
+{\mathcal L}_{\xi q}^{(1)}+{\mathcal L}_{\rm YM}^{(1)}$ refers to the 
${\cal O}(\lambda)$ suppressed terms in the SCET 
Lagrangian given in Ref.~\cite{Beneke:2002ni}. It is understood that 
the collinear fields in these terms are those of direction $i$. The 
generalization to ${\cal O}(\lambda^2)$ should be evident.

In the following, we will focus on the case in which one of the collinear 
directions carries fermion number $F=2$. The simplification of this 
choice results from the absence of a leading-power operator $J_i^{A0}$ 
(and consequently all $J_i^{An}$), since one needs two fermion fields in 
the same direction to begin with. Nevertheless, this simpler case allows 
us to display most of the features of the anomalous dimension at 
${\cal O}(\lambda^2)$. The $F=2$ operator basis at ${\cal O}(\lambda)$ 
consists of the single collinear operator 
\bea
\label{eq:Jxixi}
J^{B1}_{\chi_\alpha \chi_\beta}(t_{i_1},t_{i_2}) &=& 
\chi_{i \alpha} (t_{i_1}\nnp{i})\chi_{i\beta} (t_{i_2}\nnp{i}) \;.
\eea
We keep open the Dirac spinor indices $\alpha,\beta$, because they will in 
general be contracted with components of the $N$-jet operator from the 
other collinear directions $j\not=i$. The same rule applies to Lorentz and 
colour indices, and we only assume that the total $N$-jet operator transforms 
as a colour singlet. At  ${\cal O}(\lambda^2)$, we have 
\bea
J^{B2}_{\chi_\alpha \partial^\mu\chi_\beta} (t_{i_1},t_{i_2}) &=&  
\chi_{i\alpha }(t_{i_1}\nnp{i})i\partial_{\perp i}^\mu 
\chi_{i\beta}(t_{i_2}\nnp{i})\,,
\nn\\
J^{B2}_{\partial^\mu(\chi_\alpha \chi_\beta)}(t_{i_1},t_{i_2}) &=& 
i\partial_{\perp i}^\mu J^{B1}_{\chi_\alpha \chi_\beta}(t_{i_1},t_{i_2})\,,
\nn\\
J^{C2}_{{\cal A}^\mu \chi_\alpha \chi_\beta}(t_{i_1},t_{i_2},t_{i_3}) &=&  
{\cal A}_{\perp i}^\mu(t_{i_1}\nnp{i}) 
\chi_{i\alpha}(t_{i_2}\nnp{i})\chi_{i\beta}(t_{i_3}\nnp{i})\,.
\label{eq:Jxixi2}
\eea
We will omit the Dirac indices in the following for brevity and drop the 
direction index $i$ in the notation for the operator unless ambiguities 
can arise. The time-ordered product operators at $\order{\lambda^2}$ are
\bea
J^{T2}_{\chi \chi,\xi}(t_{i_1},t_{i_2}) &=& 
i \int d^4 x \,T\left\{J^{B1}_{\chi \chi}(t_{i_1},t_{i_2}), 
{\mathcal L}_\xi^{(1)}(x) \right\},\nn\\
J^{T2}_{\chi \chi,\xi q}(t_{i_1},t_{i_2}) &=&
i \int d^4 x \,T\left\{J^{B1}_{\chi \chi}(t_{i_1},t_{i_2}), 
{\mathcal L}^{(1)}_{\xi q}(x) \right\},\nn\\
J^{T2}_{\chi \chi, \rm YM}(t_{i_1},t_{i_2}) &=&i 
\int d^4 x \,T\left\{J^{B1}_{\chi \chi}(t_{i_1},t_{i_2}), 
{\mathcal L}^{(1)}_{\rm YM}(x) \right\}.
\label{eq:Ti}
\eea
The inclusion of these operators guarantees that the anomalous dimension 
matrix does not mix operators with different $\lambda$ scaling. Note 
that in contrast to the local current operators, the time-ordered products 
always contain the soft fields. 

\section{Anomalous dimension}
\label{sec:anomdim}

\subsection{General structure}

The operator renormalization in renormalized perturbation theory is given by
\bea
\langle {\cal O}_{P}(\{\phi_{\rm ren}\},\{g_{\rm ren}\})\rangle_{\rm ren} 
&=& \sum_Q Z_{PQ}\prod_{\phi\in Q} Z_\phi^{1/2}\prod_{g\in Q} Z_g\langle 
{\cal O}_{Q, \rm bare}(\{\phi_{\rm ren}\},\{g_{\rm ren}\})\rangle\,,
\eea
where $P, Q$ label the $N$-jet operators as well as time-ordered products of 
$N$-jet operators with insertions of power-suppressed interactions 
${\cal L}_{\textrm{SCET}}$. The 
products run over all fields and couplings that enter 
$\langle{\cal O}_Q\rangle$, respectively. We omit the argument in the 
following for brevity. At one-loop, writing 
$Z_{PQ}=\delta_{PQ}+\delta Z_{PQ}$ and demanding that the left-hand side is 
finite, implies
\be
{\rm finite} = \langle {\cal O}_{P, \rm bare}\rangle_{\textrm{1-loop}} 
+  \sum_{Q} \left[ \delta Z_{PQ} +\delta_{PQ}\left(\frac12 \sum_{\phi\in P} 
\delta Z_\phi + \sum_{g\in P} \delta Z_g\right) \right]\langle 
{\cal O}_{Q, \rm bare}\rangle_{\rm tree} \,.
\ee
For the operator basis we are interested in we need to consider also the 
continuous operator label $x=\{x_{i_k}\}$, and generalize the anomalous 
dimension to include integrations as well as summation over different types 
of operators 
\bea\label{eq:ZAB}
{\rm finite} &=& \langle J_{P}(x)\rangle_{\textrm{1-loop}} \\
  && + \sum_{Q}\int dy \left[ \delta Z_{PQ}(x,y) 
+\delta_{PQ}\delta(x-y)\left(\frac12 \sum_{\phi\in P} \delta Z_\phi 
+ \sum_{g\in P} \delta Z_g\right)\right]\langle J_Q(y)\rangle_{\rm tree} 
\,, \nn
\eea
where $\delta(x-y)\equiv \prod_{i}\prod_{k=2}^{n_i}\delta(x_{i_k}-y_{i_k})$ 
and accordingly $Z_{PQ}(x,y)=\delta_{PQ}\delta(x-y)+\delta Z_{PQ}(x,y)$. Note 
that for $n_i$ collinear building blocks in one direction we need $n_i-1$ 
integrals, because $\sum_k x_{i_k}=1$. If there is only a single building 
block for a given direction $i$, then $x_{i_1}=1$, and no integration over 
momentum fractions occurs. We use the convention that empty products are 
unity, so that the above equation covers also this case.

As discussed below, the soft loops within a single collinear direction vanish.
Therefore, we split the renormalization constant to soft and collinear 
contributions via
\be
  \delta Z_{PQ}(x,y) = \sum_{i\not= j} \delta(x-y)\delta Z_{PQ}^{s, ij}(x) 
+ \sum_i \delta^{[i]}(x-y)\delta Z_{PQ}^{c, i}(x,y)\,,
\ee
where we have used the fact that the soft loops are diagonal in $x$.
The collinear loop along direction $i$ is diagonal in the $x_{j_k}$ for 
$j\not=i$, which is reflected by  $\delta^{[i]}(x-y)\equiv 
\prod_{j\not=i}\prod_{k>1}\delta(x_{j_k}-y_{j_k})$.
This gives the $\overline{\rm MS}$ scheme renormalization conditions
\bea
0 &=& \langle J_P(x)\rangle_{\textrm{1-loop, div.}}^{\rm soft, ij} 
+ \sum_{Q}\delta Z_{PQ}^{s, ij}(x) \langle J_Q(x)\rangle_{\rm tree} \,,
\\ 
0 &=& \langle J_P(x) \rangle_{\textrm{1-loop, div.}}^{\rm coll., i} 
+ \sum_{Q}\int \prod_{k>1} dy_{i_k}  \Bigg[ \delta Z_{PQ}^{c, i}(x,y) 
\nn\\
&& {} +\delta_{PQ}\prod_{k>1} \delta(x_{i_k}-y_{i_k})
\left(\frac12 \sum_{\phi\in J_{Pi}} \delta Z_\phi + \sum_{g\in J_{Pi}} 
\delta Z_g\right) \Bigg]\langle J_Q(y)\rangle_{\rm tree} \,
\label{eq:ZABcoll},
\eea
where in the collinear part $x_{j_k}=y_{j_k}$ for $j\not=i$. In the last line 
we include only those field- and coupling renormalization factors that are 
associated to collinear building blocks of the direction $i$ (that is, 
$\frac12\delta Z_\chi=-\frac{\als C_F}{8\pi\epsilon}$ for each collinear 
fermion, and $\frac12\delta Z_A + \delta Z_{g_s}=-\frac{\als C_A}
{4\pi\epsilon}$ for each collinear gluon). We also use the notation 
$Z_{PQ}^{c, i}(x,y)=\delta_{PQ}\prod_{k>1} \delta(x_{i_k}-y_{i_k})
+\delta Z_{PQ}^{c, i}(x,y)$. 

The anomalous dimension matrix is defined by
\be
{\bf \Gamma} = - {\bf Z}^{-1} \frac{d}{d\ln\mu} {\bf Z}\,,
\ee
where we use matrix notation involving both discrete indices ($P,Q$) 
labelling the set of $N$-jet operators including open Lorentz, spinor and  
colour indices as well as continuous indices $(x,y)$ for the
collinear momentum fractions associated to each building block.

Before we proceed to discuss the details of each contribution, let us make a 
technical remark about the extraction of ultraviolet (UV) divergences. 
To compute the 
anomalous dimension we need to separate the UV and infrared (IR) poles of the amplitude. 
 In our computation of the soft and collinear contributions we assume that the 
external states have small off-shellness $p_{i_k}^2\neq 0$. This choice 
regularizes the IR divergences of the amplitude and guarantees that all the 
$1/\epsilon^n$ divergences are related to UV poles of the SCET amplitude. At 
the end of the computation, the soft and collinear part are combined and only 
then the limit $p_{i_k}^2\to 0$ can be taken. The cancellation of the 
off-shell regulator dependence serves as an additional check of our 
computation.

\subsection{Collinear part}

The collinear contribution to the anomalous dimension can be extracted by 
computing one-loop matrix elements with a collinear loop. These loops do not 
contain soft fields, and therefore it is sufficient to concentrate on
purely collinear interactions. In principle, there could be collinear 
one-loop diagrams with external soft gluons generated by the insertion of 
a power-suppressed Lagrangian interaction. The divergent part of 
any such diagram would correspond to the mixing of one of the time-ordered 
product operators into a current operator with a soft field. 
However, as shown in the previous section there are no such operators 
at ${\cal O}(\lambda^2)$ that cannot be removed by the field equations. 
It is therefore sufficient to focus on collinear loop amplitudes with 
external collinear lines only. 

Since each collinear sector is interacting 
only with itself, collinear contributions factorize into individual 
contributions from each of the collinear directions $\nnp{i}$, $i=1,\dots,N$, 
respectively.  Therefore, it is sufficient to consider only the contribution 
$J_i$ to the $N$-jet operator that contains collinear fields along the 
$\nnp{i}$-direction, while the other contributions $J_{j\not=i}$ are 
irrelevant. Moreover, in the position-space SCET formulation there are no 
purely collinear power-suppressed interactions, so the power counting of the 
collinear loop is determined solely by the operator.  We first consider the 
case of an ${\cal O}(\lambda)$ power suppressed operator $J_i$, and then 
turn to the more involved case of ${\cal O}(\lambda^2)$, where operator 
mixing occurs. We will often omit the label $i$ of the collinear quantities 
in this section for brevity, since only a single collinear direction is 
involved.

\subsubsection{Order ${\cal O}(\lambda)$}

\begin{figure}[t]
\includegraphics[width=\textwidth]{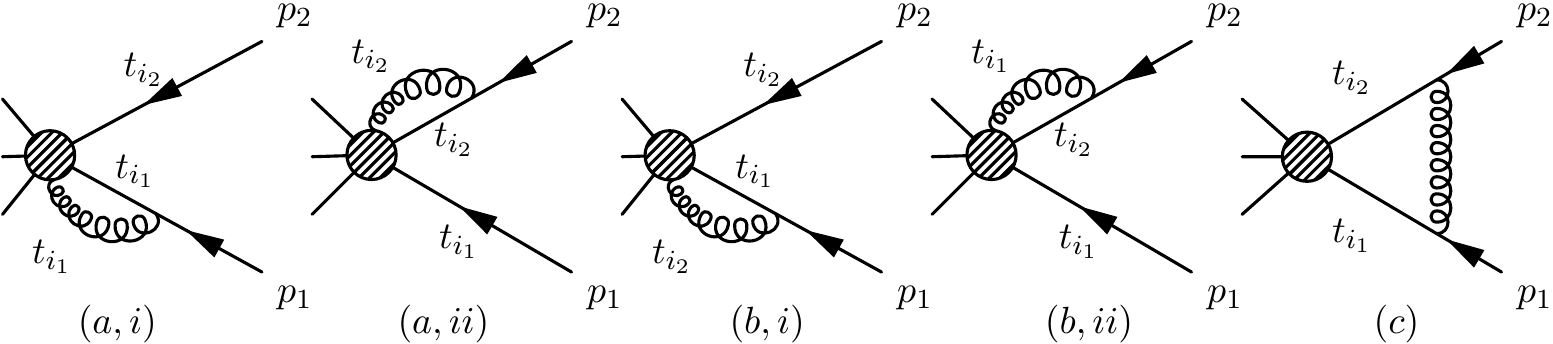}
\caption{\label{fig:Jxixi} Collinear loops contributing to the anomalous 
dimension for two fermionic building blocks in direction $\nnp{i}$. Arrows 
show the fermion flow for two outgoing antiquarks. 
}
\end{figure}

In order to extract the anomalous dimension, we consider the matrix element 
of $J^{B1}_{\chi \chi}$ defined in Eq.\,\eqref{eq:Jxixi} with two external 
fermions with external momenta $p_1$ and $p_2$. To be specific, we take the 
two fermions to be distinguishable by their flavours, which we do not show 
explicitly. The extension to identical particles will be discussed below 
Eq.~(\ref{eq:Xlocal}). We show the collinear one-loop diagrams in 
Fig.~\ref{fig:Jxixi}. The labels $t_{i_1}$ and $t_{i_2}$ indicate whether the 
corresponding line is attached to the first or second building block of 
$J^{B1}_{\chi \chi}$.

For the first two diagrams, all internal lines contributing to the collinear 
loop are attached to a single building block. In the following, we refer to 
these contributions as type-(a) loops. Since effectively only a single 
building block is involved, type-(a) loops can be inferred from the 
leading-power result. In particular, collecting the sum of the  
two type-(a) one-loop 
diagrams, the tree-level diagram, and the contributions from wave-function 
renormalization from the right-hand side of Eq.\,\eqref{eq:ZABcoll}
for the two external building blocks  of $J^{B1}_{\chi \chi}$
in a matrix element labelled with subscript (a), we find
\be
\label{eq:JxixiTypeA}
\langle \bar q(p_1)\bar q(p_2)|J^{B1}_{\chi\chi}(t_{i_1},t_{i_2})|0
\rangle_{(a)} = J_q(p_1^2)J_q(p_2^2)\langle \bar q(p_1)\bar q(p_2)|
J^{B1}_{\chi\chi}(t_{i_1},t_{i_2})|0\rangle_{\rm tree}\,,
\ee
where
\be\label{eq:Jq}
J_q(p^2) = 1 + \frac{\als C_F}{4\pi}
\left[ \frac{2}{\epsilon^2}+\frac{2}{\epsilon}
\ln\left(\frac{\mu^2}{-p^2}\right)+\frac{3}{2\epsilon}\right]
+{\cal O}(\epsilon^0)
\ee
is the leading-power collinear contribution from a single fermionic building 
block \cite{Becher:2009qa, Becher:2014oda}.

The third and fourth diagram in Fig.~\ref{fig:Jxixi} appear similar to the 
first and second one at first sight, but differ in an important respect.
Namely, the two internal lines of the collinear loop are attached to 
two \emph{different} building blocks of  $J^{B1}_{\chi \chi}$. As a 
consequence, the fractions of collinear momenta of the two lines attached to 
the operator will in general be different from the momentum fractions of the 
external lines.
We consider the operator $J^{B1}_{\chi \chi}(x)$ in Fourier space with respect to the collinear direction, where $x$ denotes the momentum fraction associated to the first building block, and correspondingly $\bar x=1-x$ for the second building block. For the external momenta, we label the collinear momentum fractions by $y=\nnp{i}p_1/(\nnp{i}p_1+\nnp{i}p_2)=\nnp{i}p_1/P$  and $\bar y=1-y$. In this notation, the tree-level diagram is in collinear momentum space given by
\bea
  \langle \bar q(p_1)\bar q(p_2)|J^{B1}_{\chi_\alpha\chi_\beta}(y)|0\rangle_{\rm tree} &=& -\delta(yP-\np p_1)\delta(\bar yP-\np p_2)v_{ \alpha}(p_1) v_{ \beta}(p_2) \, ,
\eea
where $v_{\alpha}(p) $ is the collinear spinor for the outgoing antiquark with momentum $p$ and spinor index $\alpha$.
In order to compute the one-loop matrix element in collinear momentum space, we express loop integrals $d^dl=\frac12 d\np l\, d\nm l\, d^{d-2}l_\perp$ 
in light-cone coordinates, and first perform the $\nm l$ integration by closing the contour either in the upper or lower complex plane. Then the
integration over $l_\perp$ can be performed by standard techniques, while the integration over $\np l$ is trivial and set by
the fixed value of the momentum fraction $x$ in collinear momentum space. Finally, we express the result in terms of the tree-level matrix element
by first renaming $y\to y'$, inserting $1=\int dy \delta(y-y')$, and using
\be
\label{eq:deltatrf}
\delta(yP-\np p_1)\delta(\bar yP-\np p_2)=\frac{1}{P}\delta(P-\np (p_1+p_2))
\delta(y-\np p_1/P)\,.
\ee
For example, in position space we find for the contribution from diagram $(b, i)$
\bea
\lefteqn{ \langle \bar q(p_1)\bar q(p_2)|J^{B1}_{\chi_\alpha\chi_\beta}
(t_{i_1},t_{i_2})|0\rangle_{(b,i)} 
= \tilde\mu^{4-d}\!\int\frac{d^dl}{(2\pi)^d}\,e^{i(t_{i_1}\np(p_1-l)+t_{i_2}\np(p_2+l))} } 
\nn\\
&& \times \left[\frac{i\np(l-p_1)\frac{\slashed{n}_-}{2}}{(l-p_1)^2}ig_s\nm^\mu\frac{\slashed{n}_+}{2} t^av_{\alpha}(p_1)\right] \, \frac{g_s {\np}_\mu}{\np l}t^av_{\beta}(p_2) \, \frac{-i}{l^2} 
\nn\\
&=& \frac{\als e^{\gamma_E\epsilon}\Gamma(\epsilon)}{2\pi}\,
[t^av_{\alpha}(p_1)][t^av_{\beta}(p_2)]\int_0^1dz \left(\frac{\mu^2}{-p_1^2z\bar z}\right)^\epsilon\frac{\bar z}{z} \,
  e^{i(t_{i_1}\bar z\np p_1+t_{i_2}(\np p_2+z\np p_1))}\,,\quad
\eea
where $z=\np l/\np p_1$, $\bar z=1-z$, and $l$ is the momentum of the gluon in the loop.
Fourier transforming to collinear momentum space yields a delta function that allows to trivially evaluate the $z$ integration.
Following the steps described above we obtain
\bea
\lefteqn{ \langle \bar q(p_1)\bar q(p_2)|J^{B1}_{\chi_\alpha\chi_\beta}(x)|0
\rangle_{(b,i)}  } 
\nn\\
&=& -\frac{\als e^{\gamma_E\epsilon}\Gamma(\epsilon)}{2\pi}\int_0^1dy \,
\theta(y-x)\left(\frac{\mu^2y^2}{-p_1^2x(y-x)}\right)^\epsilon\frac{x}{y(y-x)} \nn\\
&& {} \times
   {\bf T}_{i_1}\cdot{\bf T}_{i_2} \langle \bar q(p_1)\bar q(p_2)|J^{B1}_{\chi_\alpha\chi_\beta}(y)|0\rangle_{\rm tree} \nn\\
&=& -\frac{\als }{2\pi}\int_0^1dy \,\Bigg\{ \frac{1}{\epsilon}\theta(y-x)\frac{x}{y(y-x)_+} 
   -\delta(x-y)\left[\frac{1}{\epsilon^2}+\frac{1}{\epsilon}\ln\left(\frac{\mu^2 x}{-p_1^2\bar x}\right)\right]+{\cal O}(\epsilon^0)\,\Bigg\} \nn\\
&& {} \times {\bf T}_{i_1}\cdot{\bf T}_{i_2} \langle \bar q(p_1)\bar q(p_2)|J^{B1}_{\chi_\alpha\chi_\beta}(y)|0\rangle_{\rm tree}\,,
\eea
where we used colour-space operator notation for the generators, $[t^av_{\alpha}(p_1)][t^av_{\beta}(p_2)]\to {\bf T}_{i_1}\cdot{\bf T}_{i_2} v_{\alpha}(p_1)v_{\beta}(p_2)$.
Here ${\bf T}_{i_1}$ and ${\bf T}_{i_2}$ are understood to act on the fundamental colour index of the first and second building block of $J^{B1}_{\chi \chi}$,
respectively. Diagram $(b, ii)$ gives a similar result, that differs only by the replacement $x\leftrightarrow\bar x$, $y\leftrightarrow\bar y$
and $p_1^2 \leftrightarrow p_2^2$ outside of
the matrix elements. For diagram $(c)$ we find
\bea
\langle \bar q(p_1)\bar q(p_2)|J^{B1}_{\chi_\alpha\chi_\beta}(x)|0
\rangle_{(c)} &\to& -\frac{\als {\bf T}_{i_1}\cdot{\bf T}_{i_2}}
{8\pi\epsilon}\int dy\left(\theta(x-y)\frac{\bar x}{\bar y}+\theta(y-x)
\frac{x}{y}\right)
\nn\\
&&  {}\times\left(\gamma_\perp^\nu\gamma_\perp^\mu \right)_{\alpha\gamma}
\left(\gamma_{\perp\nu}\gamma_{\perp\mu}\right)_{\beta\delta} 
\langle \bar q(p_1)\bar q(p_2)|J^{B1}_{\chi_\gamma\chi_\delta}(y)|0
\rangle_{\rm tree}\,.
\quad
\eea
Note that this contribution induces a spin-dependent structure, i.e.~it is 
non-diagonal in Dirac indices.
Collecting all results, we can read off the collinear contribution to the anomalous dimension using Eq.\,\eqref{eq:ZABcoll}.
It has a diagonal part $\propto \delta(x-y)$ in collinear momentum space, and a non-diagonal part.
Using \eqref{eq:Jq} and ${\bf T}_{i_k}^2=C_F$ for quarks, we can write the anomalous dimension in the
form
\bea\label{eq:ZJxixi}
\delta Z^{c,i}_{\chi_{\alpha}\chi_{\beta},\chi_{\gamma}\chi_{\delta}}(x,y) 
&=& - \delta(x-y)\delta_{\alpha \gamma}\delta_{\beta \delta} X_{i_1i_2} + 
\frac{1}{\epsilon}\,\gamma^i_{\chi_{\alpha}\chi_{\beta},\chi_{\gamma}
\chi_{\delta}}(x,y)\,,
\eea
with
\bea
  X_{i_1i_2} &\equiv& \frac{\als}{4\pi} \Bigg\{\frac{2}{\epsilon^2}({\bf T}_{i_1}+{\bf T}_{i_2})^2 + \frac{2}{\epsilon}({\bf T}_{i_1}+{\bf T}_{i_2})\cdot\Bigg[{\bf T}_{i_1}\ln\left(\frac{\mu^2 }{-p_1^2}\right) \nn\\
  && + {\bf T}_{i_2}\ln\left(\frac{\mu^2 }{-p_2^2}\right) \Bigg] + \frac{1}{\epsilon}\left({\bf T}_{i_1}^2c_{i_1}+{\bf T}_{i_2}^2c_{i_2}\right)  \Bigg\} \,,
\eea
and
\bea\label{eq:gammaxixi}
\gamma^i_{\chi_{\alpha}\chi_{\beta},\chi_{\gamma}\chi_{\delta}}(x,y) &=& \frac{\als {\bf T}_{i_1}\cdot{\bf T}_{i_2}}{2\pi} \Bigg\{ \delta_{\alpha\gamma}\delta_{\beta\delta}   \Bigg( \theta(x-y)\left[\frac{1}{x-y}\right]_+ + \theta(y-x)\left[\frac{1}{y-x}\right]_+ \nn\\
  &&  {} - \theta(x-y)\frac{1-\frac{\bar x}{2}}{\bar y} -\theta(y-x)\frac{1-\frac{x}{2}}{y}\Bigg)\nn\\ 
  && -\frac14 \left(\sigma_\perp^{\nu\mu} \right)_{\alpha\gamma}\left(\sigma_{\perp\nu\mu}\right)_{\beta\delta}  \left(\theta(x-y)\frac{\bar x}{\bar y}+\theta(y-x)\frac{x}{y}\right) \Bigg\}\,.
\eea
Here we also expressed the Dirac gamma matrices in terms of $\sigma_{\perp }^{\mu\nu}\equiv \frac{i}{2} [\gamma_{\perp }^\mu,\gamma_{\perp }^\nu]$.
Note that the contributions from wave-function renormalization in Eq. \eqref{eq:ZABcoll} were already included in Eq. \eqref{eq:JxixiTypeA}, and are
thus contained in the diagonal part, with $c_{i_1}=c_{i_2}=3/2$ for quarks.

As mentioned above, in this work we restrict the discussion to the case of 
two-fermion operators. Detailed results for all possible contributions to the 
$N$-jet operator will be presented in a forthcoming paper.

\subsubsection{Order ${\cal O}(\lambda^2)$}\label{sec:collLambda2}

At ${\cal O}(\lambda^2)$ the three operators in Eq.\,\eqref{eq:Jxixi2} contribute, and the anomalous dimension is correspondingly given by a $3\times 3$ block matrix.
We find the following structure at one-loop, that we will derive below: 
\be\label{eq:Zmatrix}
  \delta Z^{\ci}_{PQ} = \quad 
  \begin{array}{c||cc|c}
                 & J^{B2}_{\chi\partial\chi} & J^{B2}_{\partial(\chi\chi)} & J^{C2}_{{\cal A}\chi\chi} \\ \hline\hline
  J^{B2}_{\chi\partial\chi} &\eqref{eq:ZJxidxiJxidxi}&\eqref{eq:ZJxidxiJdxixi}& \eqref{eq:ZJxidxiJAxixi}\\
  J^{B2}_{\partial(\chi\chi)} &0&\eqref{eq:ZJdxixiJdxixi}&0 \\ \hline
  J^{C2}_{{\cal A}\chi\chi} &0&0& \eqref{eq:ZJAxixiJAxixi}\\
  \end{array}
\ee
The equation numbers point to the results for the non-zero entries. Note that the operators $J_i^{T2}$ containing insertions of the power-suppressed
SCET Lagrangian contain at least one soft field and therefore do not contribute in the purely collinear sector.
We first discuss the first row $\delta Z^{c,i}_{\chi\partial\chi,Q}$, then the second $\delta Z^{c,i}_{\partial(\chi\chi),Q}$, and finally the last row $\delta Z^{c,i}_{{\cal A}\chi\chi,Q}$,
where $Q\in\{\chi\partial\chi,\partial(\chi\chi),{\cal A}\chi\chi\}$. The zero entries in the second row persist at higher orders in $\alpha_s$ (see below).

\paragraph{First row:} 
The contributions $\delta Z^{c,i}_{\chi\partial\chi,\chi\partial\chi}$ and $\delta Z^{c,i}_{\chi\partial\chi,\partial(\chi\chi)}$ can be extracted by computing the 
matrix element $\langle \bar q(p_1)\bar q(p_2)|J^{B2}_{\chi\partial\chi}(x)|0\rangle$ at one-loop, involving diagrams as in Fig.~\ref{fig:Jxixi}.
The additional $\partial_\perp$ derivative leads to an extra power of the loop momentum in the numerator, which yields a more involved structure
of divergences compared to ${\cal O}(\lambda)$.
The divergent part can be expressed in terms of the two tree-level contributions $\langle \bar q(p_1)\bar q(p_2)|J^{B2}_{\chi\partial\chi}(y)|0\rangle_{\rm tree}$
and $\langle \bar q(p_1)\bar q(p_2)|J^{B2}_{\partial(\chi\chi)}(y)|0\rangle_{\rm tree}$. The coefficients yield the corresponding anomalous dimensions, and we find
\bea\label{eq:ZJxidxiJxidxi}
\delta Z^{c,i}_{\chi_{\alpha}\partial^\mu\chi_{\beta}, \chi_{\alpha'}
\partial^\sigma\chi_{\beta'}}(x,y) 
&=& - \delta(x-y)\delta_{\alpha\alpha'}\delta_{\beta\beta'}\,
g_\perp^{\mu\sigma} 
X_{i_1i_2} + 
\frac{1}{\epsilon}\,\gamma^i_{\chi_\alpha\partial^\mu\chi_\beta, 
\chi_{\alpha'}\partial^\sigma\chi_{\beta'}}(x,y)\,, \quad\\
\delta Z^{c,i}_{\chi_\alpha\partial^\mu\chi_\beta,\partial^\sigma(\chi_{
\alpha'}\chi_{\beta'})}(x,y) 
&=&  \frac{1}{\epsilon}\,\gamma^i_{\chi_\alpha\partial^\mu\chi_\beta, 
\partial^\sigma(\chi_{\alpha'}\chi_{\beta'})}(x,y) \,, 
\label{eq:ZJxidxiJdxixi}
\eea
with
\bea\label{eq:gammaJxidxiJxidxi}
  \lefteqn{ \gamma^i_{\chi_\alpha\partial^\mu\chi_\beta, \chi_{\alpha'}\partial^\sigma\chi_{\beta'}}(x,y) \ }\nn\\
  &=&  \frac{\als {\bf T}_{i_1}\cdot{\bf T}_{i_2}}{2\pi}\Bigg\{\delta_{\alpha\alpha'}\delta_{\beta\beta'}g_\perp^{\mu\sigma}  \Bigg( \theta(x-y)\left[\frac{1}{x-y}\right]_+ + \theta(y-x)\left[\frac{1}{y-x}\right]_+ \nn\\
  && {} - \theta(x-y)\frac{\bar x+\bar y}{\bar y^2} -\theta(y-x)\frac{x+y}{y^2}\Bigg)
  +\frac14 M_{\chi_\alpha\partial^\mu\chi_\beta, \chi_{\alpha'}\partial^\sigma\chi_{\beta'}}(x,y)\Bigg\}\,,\nn\\
 \lefteqn{ \gamma^i_{\chi_\alpha\partial^\mu\chi_\beta, \partial^\sigma(\chi_{\alpha'}\chi_{\beta'})}(x,y) \ }\nn\\
 &=&    \frac{\als {\bf T}_{i_1}\cdot{\bf T}_{i_2} }{2\pi}\left( \delta_{\alpha\alpha'}\delta_{\beta\beta'}g_\perp^{\mu\sigma}\theta(y-x)\frac{x}{y^2}
 +\frac14 M_{\chi_\alpha\partial^\mu\chi_\beta, \partial^\sigma(\chi_{\alpha'}\chi_{\beta'})}(x,y)\right)\,.
\eea
The last terms in each expression arise from diagram (c) and are given in 
App.~\ref{sec:aux}.

Let us now turn to $Z^{c,i}_{\chi\partial\chi,{\cal A}\chi\chi}$, which describes the mixing of B- into C-type operators.
To extract this contribution we compute the matrix element $\langle g(q)\bar q(p_1)\bar q(p_2)|J^{B2}_{\chi\partial^\mu\chi}|0\rangle$
at one-loop involving a gluon and two antiquarks. To determine the mixing with $J^{C2}_{{\cal A}\chi\chi}$ it is sufficient
to consider a configuration where the gluon has only $\perp$ polarization, and the external
 momenta of all particles have vanishing $\perp$ component.

For loops that consist of two internal lines that are both attached to the same building block of the operator $J^{B2}_{\chi\partial^\mu\chi}$
(called type-(a) loops above) one of the collinear building blocks, that is not contributing to the loop, acts as a `spectator',
i.e. the matrix element factorizes,
\bea
  \lefteqn{\langle g_a(q)\bar q(p_1)\bar q(p_2)|J^{B2}_{\chi^\alpha\partial^\mu\chi^\beta}|0\rangle_{(a)} }\nn\\
  &=& \langle g_a(q)\bar q(p_1)| \chi(t_{i_1}\np)|0\rangle_{(a)}(-p_{2\perp}^\mu)\langle \bar q(p_2)|\chi(t_{i_2}\np)|0\rangle_{\rm tree}\nn\\
  && {} + \langle \bar q(p_1)| \chi (t_{i_1}\np)|0\rangle_{\rm tree}(-p_2-q)_{\perp}^\mu\langle g_a(q)\bar q(p_2)|\chi(t_{i_2}\np)|0\rangle_{(a)}\,,
\eea
where we have also used that the $\perp$ derivative acting on the second 
building block gives a simple factor of total momentum both at the tree- and 
loop-level. All contributions of type-(a) are therefore zero for vanishing 
external $\perp$ momenta.

\begin{figure}[t]
\begin{center}
\includegraphics[width=0.9\textwidth]{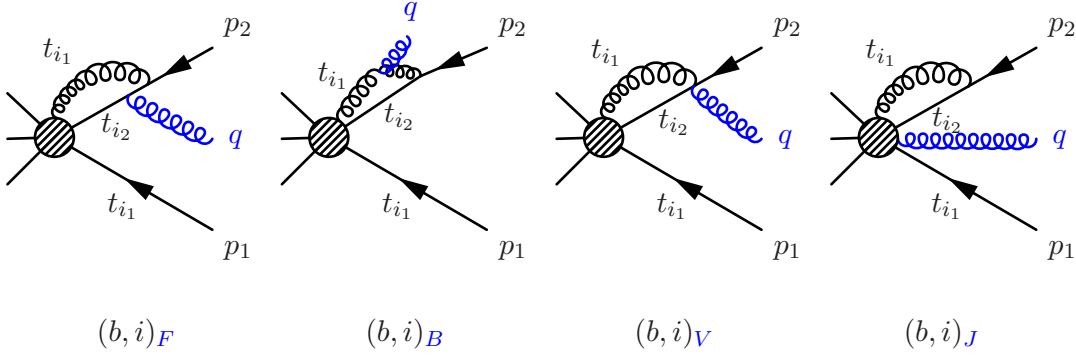}
\end{center}
\caption{\label{fig:BCmixing_example} Examples for the four possibilities of 
adding an extra collinear emission (indicated by the blue line) to a diagram 
with two fermion lines (chosen to be diagram $(b,i)$ from 
Fig.~\ref{fig:Jxixi}. for illustration).}
\end{figure}

\begin{figure}[t]
\begin{center}
\includegraphics[width=0.66\textwidth]{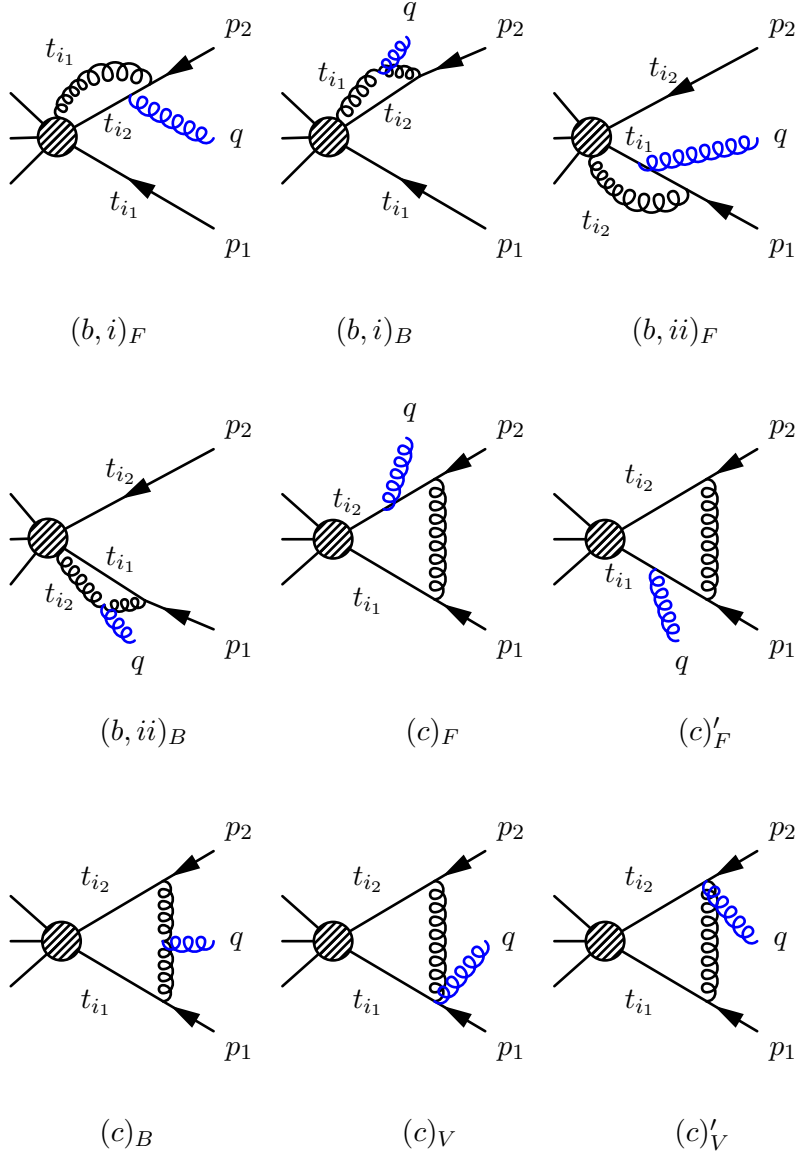}
\end{center}
\caption{\label{fig:BCmixing} 
Collinear loops contributing to the anomalous dimension 
$Z^{c,i}_{\chi\partial\chi,{\cal A}\chi\chi}$, that describes
mixing of B- into C-type operators. Arrows show the fermion flow for two 
outgoing antiquarks.}
\end{figure}

Therefore, we can focus on loops that connect the two building blocks.
They are obtained from the one-loop diagrams for a quark-quark matrix element 
shown in Fig.~\ref{fig:Jxixi} (specifically from diagrams $(b, i)$, $(b,ii)$ 
and $(c)$)
with the additional emission of the gluon off either an internal fermion line (subscript $F$), internal boson (i.e. gluon) line ($B$), vertex ($V$), or directly
from the operator ($J$). These four possibilities are illustrated in Fig.~\ref{fig:BCmixing_example} for diagram $(b, i)$.
The case ($J$) is only possible if the gluon is attached to a Wilson line, and therefore this contribution vanishes for $\perp$ polarization. 
Analogous arguments hold for $(b,ii)$ and $(c)$.
Similarly, the contributions $(b, i)_V$, $(b, ii)_V$ are zero, because the internal gluon 
is in this case connected to a Wilson line, and the four-point vertex connecting two collinear gluons and two collinear quarks vanishes when contracted with $\np^\mu$.
Finally, there could be a contribution from one-particle reducible (1PR) diagrams for which the 1PR propagator is canceled by a corresponding momentum-squared suppression of the
loop diagram. However, it turns out that there are no such contributions because the
vertex for radiating off a $\perp$ polarized gluon from a quark line with momentum $p$ vanishes for $p_\perp=0$.
In summary, all relevant loop diagrams are shown in Fig.\,\ref{fig:BCmixing}.

The computation of the one-loop diagrams is straightforward and we find the 
result
\bea\label{eq:ZJxidxiJAxixi}
  \lefteqn{ \delta Z^{c,i}_{\chi^s_\alpha\partial^\mu\chi^t_\beta,{\cal A}^{\nu a}\chi^k_{\alpha'}\chi^l_{\beta'}}(x,y_1,y_2)\ }\nn\\
  &=& \frac{\als}{8\pi\epsilon}\Big\{
  -i f^{abc} t^{c}_{sk} t^b_{tl} K_{1,\alpha\alpha'\beta\beta'}^{ \mu\nu}(x,y_2,y_3) \nn\\
  && + (t^at^{b})_{sk} t^b_{tl}K_{2, \alpha\alpha'\beta\beta'}^{\mu\nu}(x,y_1,y_2) 
  - (t^at^{b})_{tl} t^b_{sk} K_{2,\beta\beta'\alpha\alpha'}^{\mu\nu }(\bar x,y_1,y_3) \Big\}\nn\\
  &=& \frac{1}{\epsilon}\, \gamma^{i}_{\chi^s_\alpha\partial^\mu\chi^t_\beta,{\cal A}^{\nu a}\chi^k_{\alpha'}\chi^l_{\beta'}}(x,y_1,y_2)\,,
\eea
where we made explicit colour indices for clarity. The $y_k$ denote the 
collinear momentum fractions for $J^{C2}_{{\cal A}\chi\chi}$ with 
$y_1+y_2+y_3=1$, and $y_1$ corresponds to the gluonic building block 
${\cal A}$. The kernels $K$ are defined in App.~\ref{sec:aux}.
In colour-space notation
\bea\label{eq:gammaJxidxiJAxixi}
  \lefteqn{ \gamma^{i}_{\chi_\alpha\partial^\mu\chi_\beta,{\cal A}^{\nu}\chi_{\alpha'}\chi_{\beta'}}(x,y_1,y_2) 
  \ =\ \frac{\als}{8\pi}\Big\{
{\bf T}_{i_1} \times {\bf T}_{i_2} K_{1, \alpha\alpha'\beta\beta'}^{\mu\nu}(x,y_2,y_3)  }\nn\\
   && - {\bf T}_{i_1}({\bf T}_{i_1}\cdot{\bf T}_{i_2}) K_{2,\alpha\alpha'\beta\beta'}^{\mu\nu}(x,y_1,y_2) 
  + {\bf T}_{i_2}({\bf T}_{i_2}\cdot{\bf T}_{i_1}) K_{2,\beta\beta'\alpha\alpha'}^{\mu\nu}(\bar x,y_1,y_3) \Big\}\,,
\eea
where we defined a cross product via $({\bf T}_{i_1} \times {\bf T}_{i_2})^a\equiv i f^{abc}{\bf T}_{i_1}^b{\bf T}_{i_2}^c$, and 
the subscripts refer to the first and second fermionic building block, respectively. In addition, we leave implicit the open adjoint index 
of the colour-space operators, which generates the additional colour label required for the gluonic building block of ${\cal A}\chi\chi$.

\paragraph{Second row:} 
Matrix elements of the operator $J^{B2}_{\partial(\chi \chi)}$ can be trivially
related to those of $J^{B2}_{\chi \chi}$, because the total derivative factors out of any loop diagram.
Therefore, we can infer the corresponding entries in the anomalous dimension matrix from the ${\cal O}(\lambda)$ result,
\bea\label{eq:ZJdxixiJdxixi}
  \delta Z^{c,i}_{\partial^\mu(\chi_\alpha\chi_\beta),\partial^\nu(\chi_\gamma\chi_\delta)}(x,y) &=& g_\perp^{\mu\nu} \delta Z^{c,i}_{\chi_\alpha\chi_\beta,\chi_\gamma\chi_\delta}(x,y), \nn\\
  \delta Z^{c,i}_{\partial(\chi\chi),Q} &=& 0 \quad (Q=\chi\partial\chi,{\cal A}\chi\chi)\;. 
\eea
The last line follows from the equality 
\be
\langle g_a(q)\bar q(p_1)\bar q(p_2)| 
J^{B2}_{\partial^\mu(\chi\chi)}|0\rangle 
= -(q+p_1+p_2)_\perp^\mu\,\langle g_a(q)\bar q(p_1)\bar q(p_2)|
J^{B1}_{\chi\chi}|0\rangle
\label{eq:JB2}
\ee
at any loop order together with the first line of Eq.~(\ref{eq:ZJdxixiJdxixi}). Since the ${\cal O}(\lambda)$ matrix element on the right-hand side is rendered finite by the $\delta Z^{c,i}_{\chi\chi,\chi\chi}$ counterterm, it is not necessary to introduce new counterterms to renormalize the left-hand side at ${\cal O}(\lambda^2)$. We have checked this explicitly by computing the left-hand side of Eq.~(\ref{eq:JB2}) at one loop. 

\paragraph{Third row:} 
For C-type operators with three collinear building blocks the one-loop anomalous dimension can be inferred from operators involving only two collinear building blocks.
The reason is that at one-loop, at most two building blocks can be connected to the loop, while the third one acts as a spectator.

In particular, type-(a) loops operate on each building block separately, and therefore give the same result as
at leading power (when including also coupling and wavefunction renormalization, as above)
\bea
  \lefteqn{\langle g_{a}(p_1)\bar q(p_2)\bar q(p_3)| J^{C2}_{{\cal A}^\mu\chi\chi}(x_1,x_2) |  0\rangle_{(a)}}\nn\\
  &=& J_g(p_1^2)J_q(p_2^2)J_q(p_3^2)\langle g_{a}(p_1)\bar q(p_2)\bar q(p_3)| J^{C2}_{{\cal A}^\mu\chi\chi}(x_1,x_2) |  0\rangle_{\rm tree}\,.
\eea
The expression for $J_q(p^2)$ is given in Eq.~(\ref{eq:Jq}), and $J_g(p^2)$ 
is given by the same expression with $C_F\to C_A$, $3/(2\epsilon)\to 0$. 

All other loops connect two building blocks. There are three possibilities to select a pair. For each pair, the computation is analogous to
the corresponding case where the third collinear building block is absent.
Therefore, we can obtain the anomalous dimension by rescaling the corresponding momentum fractions. For example, for the case where the loop connects the
second and third building block (indicated by the subscript $23$), the contribution to the anomalous dimension is related 
to  the ${\cal O}(\lambda)$ result from Eq. (\ref{eq:ZJxixi}),
\be\label{eq:23contribution}
 \left. Z^{c,i}_{{\cal A}^\mu\chi\chi,{\cal A}^\rho\chi\chi}(x_1,x_2,y_1,y_2)\right|_{23} 
  = \frac{1}{1-y_1}\delta(x_1-y_1)g_\perp^{\mu\rho} Z^{c,i}_{\chi\chi,\chi\chi}(x,y)\,,
\ee
with $x=x_2/(x_2+x_3)=x_2/(1-x_1)$ and $y=y_2/(y_2+y_3)=y_2/(1-y_1)$. The momentum fractions in the first building block are not affected by the 
loop, and therefore identical, leading to the $\delta(x_1-y_1)$, and a similar argument applies to the Lorentz indices leading to $g_\perp^{\mu\rho}$.
The prefactor is due to the Jacobian\footnote{\label{foot:norm} This can be seen by writing the corresponding delta functions in the
tree-level matrix element in the form $\delta(y_1P-\np p_1)\delta(y_2P-\np p_2)\delta(y_3P-\np p_3)
=\delta(y_1P-\np p_1)\delta(yP_{23}-\np p_2)\delta(\bar yP_{23}-\np p_3)$ where $P_{23}\equiv (1-y_1)P=(1-x_1)P$ is the collinear momentum
of the two building blocks that are connected by the loop. Then the product $\delta(yP_{23}-\np p_2)\delta(\bar yP_{23}-\np p_3)$
has the same form as for the case with only two building blocks (except that $P\to P_{23}$). The remaining factor $\delta(y_1P-\np p_1)$
is not affected by the loop integration, and therefore the same for the one-loop and tree-level matrix elements, leading to $\delta(x_1-y_1)$.
Therefore the only re-scaling factor is the Jacobian obtained from the change of integration measure $\left. Z^{\ci}_{{\cal A}^\mu\chi\chi,{\cal A}^\rho\chi\chi}(x_1,x_2,y_1,y_2)\right|_{23}dy_2dy_1 = \delta(x_1-y_1)g_\perp^{\mu\rho}dy_1 \times Z^{\ci}_{\chi\chi,\chi\chi}(x,y)dy$. For example, the Jacobian ensures that the `diagonal' contributions to $Z^{\ci}_{\chi\chi,\chi\chi}(x,y)$ have the correct
normalization, because $\delta(x_1-y_1)\delta(x-y)=(1-y_1)\delta(x_2-y_2)\delta(x_1-y_1)$.
Note also that the anomalous dimension does not explicitly depend on the total collinear momentum $P$ in the direction $\nnp{i}$ under consideration.
} $dy/dy_2=1/(1-y_1)$.

To obtain the full anomalous dimension we need to sum over the three pairs of 
collinear building blocks, $13$, $23$, $12$.
Note that the anomalous dimension on the right-hand side of Eq. (\ref{eq:23contribution}) captures also the contributions from type-(a) loops attached to either the
second or the third building block. This will also be the case for the $23$ and $12$ contributions, such that the type-(a) loops are counted twice.
We therefore need to subtract them once to obtain the correct result. In addition each term contains the tree-level contribution, which we need to
subtract twice. Altogether,
\bea\label{eq:ZJAxixiJAxixi}
 \lefteqn{Z^{c,i}_{{\cal A}^\mu\chi_\alpha\chi_\beta,{\cal A}^\nu\chi_{\alpha'}\chi_{\beta'}}(x_1,x_2,y_1,y_2) }\nn\\
 &=& \frac{1}{1-y_2}\delta(x_2-y_2)\delta_{\beta\beta'} Z^{\ci}_{{\cal A}^\mu\chi_\alpha,{\cal A}^\nu\chi_{\alpha'}}\left(\frac{x_1}{1-x_2},\frac{y_1}{1-y_2}\right) \nn\\
 && + \frac{1}{1-y_1}\delta(x_1-y_1)g_\perp^{\mu\nu} Z^{\ci}_{\chi_\alpha\chi_\beta,\chi_{\alpha'}\chi_{\beta'}}\left(\frac{x_2}{1-x_1},\frac{y_2}{1-y_1}\right)  \nn\\
 && + \frac{1}{1-y_3}\delta(x_3-y_3)\delta_{\alpha\alpha'} Z^{\ci}_{{\cal A}^\mu\chi_\beta,{\cal A}^\nu\chi_{\beta'}}\left(\frac{x_1}{1-x_3},\frac{y_1}{1-y_3}\right)   \nn\\
 && - [1+J_g(p_1^2)^{-1}J_q(p_2^2)^{-1}J_q(p_3^2)^{-1}]\delta(x_1-y_1)\delta(x_2-y_2)\delta_{\alpha\alpha'} \delta_{\beta\beta'}g_\perp^{\mu\nu}\;.
\eea
The last line contains the subtractions accounting for the over-counting (see Footnote \ref{foot:norm} for the normalization). The anomalous dimension $Z^{c,i}_{{\cal A}\chi,{\cal A}\chi}$ is given in App.~\ref{sec:aux} (see also 
Refs.~\cite{Hill:2004if, Beneke:2005gs}).
Notice that the above equation is valid only up to one-loop.
At higher loops, the three building blocks may be connected together.
Eq.\,\eqref{eq:ZJAxixiJAxixi} can be brought into the form
\bea
\delta Z^{c,i}_{{\cal A}^\mu\chi_\alpha\chi_\beta,{\cal A}^\nu\chi_{\alpha'}\chi_{\beta'}}(x_1,x_2,y_1,y_2) &=&
  - \delta_{\alpha\alpha'} \delta_{\beta\beta'}g_\perp^{\mu\nu}\delta(x_1-y_1)\delta(x_2-y_2) X_{i_1i_2i_3} 
\nn\\
&& {} + 
  \frac{1}{\epsilon}\gamma^i_{{\cal A}^\mu\chi_\alpha\chi_\beta,{\cal A}^\nu\chi_{\alpha'}\chi_{\beta'}}\,,
\eea
where
\bea
  X_{i_1i_2i_3} 
 &=& \frac{\alpha_s}{4\pi} \,\Bigg\{\frac{2}{\epsilon^2}\,({\bf T}_{i_1}+{\bf T}_{i_2}+{\bf T}_{i_3})^2 
 + \frac{2}{\epsilon}({\bf T}_{i_1}+{\bf T}_{i_2}+{\bf T}_{i_3})\cdot\Bigg[{\bf T}_{i_1}\ln\left(\frac{\mu^2 }{-p_1^2}\right) \nn\\
  && + \,{\bf T}_{i_2}\ln\left(\frac{\mu^2 }{-p_2^2}\right) + {\bf T}_{i_3}\ln\left(\frac{\mu^2 }{-p_3^2}\right) 
  \Bigg]+\frac{1}{\epsilon}\left({\bf T}_{i_1}^2c_{i_1}+{\bf T}_{i_2}^2c_{i_2}+{\bf T}_{i_3}^2c_{i_3}\right)\Bigg\} \,, \nn
\eea
with ${\bf T}_{i_1}^2=C_A$ and $c_{i_1}=0$ for the gluonic building block and ${\bf T}_{i_2}^2={\bf T}_{i_3}^2=C_F$, $c_{i_2}=c_{i_3}=3/2$
for the fermionic building blocks. The non-diagonal part is given by
\bea\label{eq:gammaJAxixiJAxixi}
 \lefteqn{\gamma^i_{{\cal A}^\mu\chi_\alpha\chi_\beta,{\cal A}^\nu\chi_{\alpha'}\chi_{\beta'}}(x_1,x_2,y_1,y_2) }\nn\\
 &=& \frac{1}{1-y_2}\delta(x_2-y_2)\delta_{\beta\beta'} \gamma^i_{{\cal A}^\mu\chi_\alpha,{\cal A}^\nu\chi_{\alpha'}}\left(\frac{x_1}{1-x_2},\frac{y_1}{1-y_2}\right) \nn\\
 && + \frac{1}{1-y_1}\delta(x_1-y_1)g_\perp^{\mu\nu} \gamma^i_{\chi_\alpha\chi_\beta,\chi_{\alpha'}\chi_{\beta'}}\left(\frac{x_2}{1-x_1},\frac{y_2}{1-y_1}\right)  \nn\\
 && + \frac{1}{1-y_3}\delta(x_3-y_3)\delta_{\alpha\alpha'} \gamma^i_{{\cal A}^\mu\chi_\beta,{\cal A}^\nu\chi_{\beta'}}\left(\frac{x_1}{1-x_3},\frac{y_1}{1-y_3}\right).
\eea
In addition, there is no mixing with operators with two building blocks, inherited from $\delta Z^{c,i}_{{\cal A}\chi,\partial\chi}=0$ at ${\cal O}(\lambda)$ 
(see App.~\ref{sec:aux}), that is, 
\be\label{eq:ZJAxixiJB}
\delta Z^{\ci}_{{\cal A}\chi\chi, Q} = 0 \quad 
(Q=\chi\partial\chi,\partial(\chi\chi))\,.
\ee

\subsubsection{General structure of the collinear anomalous dimension}

The previous findings suggest a general structure for the collinear 
contributions to the anomalous dimension. We can write schematically for the 
contribution from collinear direction $i$ with $n_i$ building blocks 
($n_i=1,2,3$ for A-, B- ,C-type operators, respectively),
\be
\label{eq:anocoll}
\delta Z^{c,i}_{PQ}(x,y) = -\delta_{PQ}\prod_k\delta(x_{i_k}-y_{i_k})
X_{i_1\dots i_{n_i}} + \frac{1}{\epsilon}\,\gamma^{i}_{PQ}(x,y)\,,
\ee
where the first term is the diagonal contribution, $\delta_{PQ}$ is non-zero 
for identical operators $P=Q$ and then stands for the product of 
$\delta_{\alpha\beta}$ for Dirac and $g_\perp^{\mu\nu}$ for Lorentz indices, 
$x_{i_k}$ and $y_{i_k}$ denote the collinear momentum fractions in direction 
$i$ for the building blocks $k=1,\dots,n_i$, and $\gamma^i_{PQ}(x,y)$ 
encapsulates the non-diagonal contribution. Here $x$ and $y$ denote the 
vectors of momentum fractions as introduced above.

The non-diagonal contributions in general encapsulate rather lengthy results 
that depend on the Lorentz structure and on momentum fractions in a generic 
way. The diagonal contribution can be summarized in a universal way,
\be\label{eq:Xlocal}
  X_{i_1\dots i_{n_i}} = \frac{\als}{4\pi} \sum_{l,k=1}^{n_i} {\bf T}_{i_l}\cdot{\bf T}_{i_k} \left\{\frac{2}{\epsilon^2} +\frac{2}{\epsilon}\ln\left(\frac{\mu^2 }{-p_{i_k}^2}\right)+\delta_{lk}\frac{c_{i_k}}{\epsilon}\right\}\,,
\ee
where $c_{i_k}=3/2$ for fermionic building blocks, and $c_{i_k}=0$ for gluonic building blocks.
For clarity we added an additional label to the off-shell regulator $p_{i_k}^2$ for the collinear direction it corresponds to.

So far we assumed that the two fermionic building blocks considered above have different flavours.
It is straightforward to generalize the result in Eq.\,(\ref{eq:anocoll}) to the case of identical building blocks, which is relevant e.g. for
quarks of identical flavour or when considering operators with more than one gluonic building block. 
For gluons (quarks), one has to symmetrize (anti-symmetrize) the anomalous dimension with respect to exchanging them (including a factor $1/N_s$ where $N_s$ is the number of terms).\footnote{In this case, the association of the external momentum with the collinear building block is not unique; however, they appear then
only in a symmetric form (e.g. $\ln(p_{i_1}^2)+\ln(p_{i_2}^2)$) such that there is no ambiguity.} 
Moreover, if more than one $\perp$ derivative acts on the same building block at $O(\lambda^2)$, 
the corresponding Lorentz indices need to be symmetrized too.

The final result for the collinear contribution to the anomalous dimension is obtained by
adding together the collinear contributions from all directions, which gives an additional sum over $i$,
\bea\label{eq:Zcoll}
  \delta Z^c_{PQ}(x,y) &=& \sum_{i=1}^N \delta^{[i]}(x-y) \delta Z^{c,i}_{PQ}(x,y) \nn\\
  &=& -\delta_{PQ}\delta(x-y)\sum_i X_{i_1\dots i_{n_i}} + \sum_i\delta^{[i]}(x-y)\frac{\gamma^i_{PQ}(x,y)}{\epsilon}\,,
\eea
where we used that $\delta^{[i]}(x-y)\prod_{k>1}\delta(x_{i_k}-y_{i_k})=\delta(x-y)$ in the compact vector notation introduced above.
This result is consistent with all individual results obtained above, for the fermion number two case.
We checked that the diagonal contributions are in accord with Eq. \eqref{eq:Zcoll} also for fermion number one and zero up to ${\cal O}(\lambda^2)$.


\subsection{Soft part}

The soft fields mediate interactions between collinear fields in different 
directions. Here, we need to consider two types of contributions: first, soft 
loops with leading-power interactions, for which the power suppression arises purely from the $N$-jet operator, giving rise to current-current mixing. Second, soft loops containing insertions of the power-suppressed contributions to the SCET Lagrangian that describe subleading soft-collinear interactions. They give rise to operator mixing involving $J_i^{T2}$ operators featuring time-ordered products, see~Eq.~(\ref{eq:Ti}). This approach helps to keep the power-counting manifest and ensures that the anomalous dimension does not mix operators with different powers of $\lambda$. Because the leading two-fermion operator is $\order{\lambda}$, in this work we need to consider only a single insertion of the subleading interaction. The leading-power interaction between soft gluons and collinear particles can be used any number of times when constructing the amplitude. 

\subsubsection{Currents}
\begin{figure}[t]
\begin{center}
\includegraphics[scale=1.0]{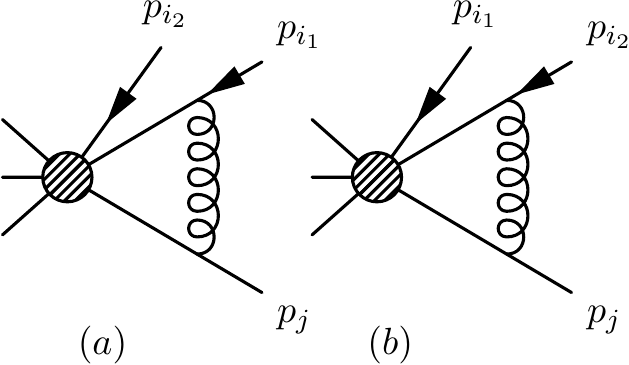}
\end{center}
\caption{\label{fig:SoftLP} The leading power diagrams with a soft-gluon 
exchange. The $j$-direction parton is either a (anti)quark or a gluon created 
by either 
A0 or A1 current. In the two-fermion sector, the current can be either B1 or 
B2.}
\end{figure}

For the current-current mixing, the 
soft loops within a single collinear sector vanish to all orders in $\als$ 
because the leading-power interaction contains only a single component of the 
soft field, $\nnm{i}\Aus $. 
Hence, to determine the soft part of the anomalous dimension we only need to 
consider soft loops connecting different collinear sectors. At the one-loop level, only two different collinear directions can be connected by a soft loop. The result is then given as a sum of all possible pairings of fields belonging to different directions. For the two-fermion operator, the relevant diagrams are presented in Fig.~\ref{fig:SoftLP}. 
The parton belonging to the $j$ direction can be either a 
(anti)quark or a gluon. 

The divergent part of the diagrams shown in Fig.~\ref{fig:SoftLP} with soft 
loops and leading power interaction is 
\begin{equation}\label{eq:Zsoft}
\delta Z_{PQ}^{s, ij}(x) =- \delta_{PQ}\,\frac{\als}{4\pi} \sum_{l=1}^{n_{i}}\sum_{k=1}^{n_{j}}\frac{\mathbf{T}_{i_l}\cdot\mathbf{T}_{j_k}}{2}\left[\frac{2}{\epsilon^{2}}+\frac{2}{\epsilon}\ln\left(\frac{-\mu^{2}  x_{i_l}x_{j_k} s_{ij}}{p_{i_l}^{2}p_{j_k}^{2}}\right)\right]\,,
\end{equation}
where $s_{ij} =\frac{1}{2} (\nnm{i} \cdot\nnm{j})  P_{i}  P_{j}$ depends only on the total collinear momentum in the directions connected by the soft loop.

The colour-space formalism reveals the universal form of the soft factor. The result in Eq.~(\ref{eq:Zsoft}) holds for gluons as well as for quarks. The soft factor depends only on the colour charge of the collinear particle but not on its spin. When there are identical partons within one collinear direction, the result should by symmetrized as in the collinear case.

The renormalization factor for the subleading currents with extra $\perp$ derivatives acting on the collinear fields is also given by Eq.\,(\ref{eq:Zsoft}). In the soft-collinear vertices, only the $\nm$ component of the momentum is conserved. The other components are conserved only within the collinear sector as dictated by the SCET multipole expansion of the soft fields. In the $\perp$ direction, the soft field wave-length is much larger than the size of typical fluctuations of the collinear field. As a result, the soft field is insensitive to the $\perp$ momentum of the collinear fields. 
Hence, the extra momentum factor in the $N$-jet operator Feynman rule that comes from the  $\perp$ derivative does not affect the computation of the soft loop. 

To summarize, the soft counterterm for the subleading local operators is universal, diagonal and given by Eq.~(\ref{eq:Zsoft}). This fact is easily understood by application of the soft decoupling transformation. The collinear fields can be redefined to remove the leading-power soft interactions from the SCET 
Lagrangian~\cite{Bauer:2001yt}. For example, for the fermion fields we define
\begin{equation}\label{eq:DecTran}
\chi(\nnp{i} t_{i_k}) =  Y_i(0) \chi^{(0)}(\nnp{i} t_{i_k}),
\quad
Y_i^\dagger(x) \equiv \mathbf{P}\exp\left[{ig_s \int_0^{\infty} ds  
\nnm{i} \Aus(x + \nnm{i} s)} \right]\,.
\end{equation}
The fields building the $N$-jet operator are evaluated at $\nnp{i} t_{i_k}$ so the decoupling transformation commutes with the derivative $\partial_{\perp i}$. The $N$-jet operator at $\order{\lambda}$ factorizes into a product of collinear fields $\chi^{(0)}$ that do not interact with the soft fields and a product of soft Wilson lines. Hence, the universality of the Eq.~(\ref{eq:Zsoft}) is a consequence of the standard eikonal approximation for the 
leading-power soft gluon coupling.

\subsubsection{Time-ordered products}

The decoupling transformation in Eq.~(\ref{eq:DecTran}) does not remove the soft fields from the non-local time-ordered product operators. In this case, it is necessary to compute the soft loops explicitly. To obtain non-zero mixing into local operators we compute diagrams where the soft field from the Lagrangian insertion appears as an internal line. Non-zero mixing can occur only between operators with identical quantum numbers, and as the local currents do not contain soft fields, only these diagrams can induce mixing into local operators. Nevertheless, we checked that the one-loop amplitudes with one external soft gluon are indeed finite after combining the soft and collinear loop contributions. 

Consider first the $J^{T2}_{\chi_\alpha \chi_\beta,\xi q}$ operator. Since there is no leading-power interaction between soft quarks and collinear partons it is impossible to form a soft loop and remove the soft quark field. Therefore, no mixing into any of the local operators is allowed for this operator. 

The operator $J^{T2}_{\chi_\alpha \chi_\beta, \rm YM}$ can form a non-vanishing contraction only with the gluon fields contained in the Wilson lines that accompany the quarks. Choosing the light-cone gauge we immediately see that this operator does not mix into any of the local operators.

\begin{figure}[t]
\begin{center}
\includegraphics[width=0.78\textwidth]{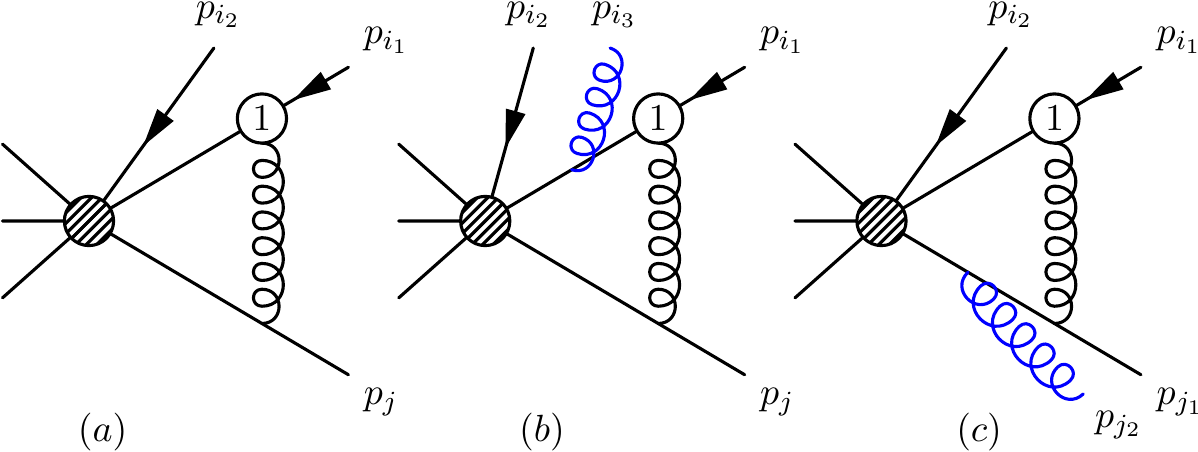}
\end{center}
\caption{\label{fig:SoftNLP} Sample  diagrams contributing to mixing of time-ordered product into power-suppressed local operators. The circle denotes the $\order{\lambda}$ SCET Lagrangian insertion. Diagram (a) contributes to mixing into $N$-jet operator with B2-type currents; the diagram (b) can induce mixing into C2-type currents and the diagram (c) can generate mixing into an $N$-jet operator containing two different B1-type operators.    }
\end{figure}

Finally, we investigate possible mixing of the time-ordered product containing $\mathcal{L}_{\xi}^{(1)}$.  The $\order{\lambda}$ Lagrangian $\mathcal{L}_{\xi}^{(1)}$ contains interactions  with  the $\perp$ and $\nm$ components of the soft field, thus it is not possible to form a contraction with the leading power soft-collinear interaction in the same collinear direction. Hence, just like in the case of local operators, the soft loops for the time-ordered product at $\order{\lambda}$ vanish within a single collinear sector. 
The soft loops connecting the time-ordered product with a different collinear direction are shown in Fig.~\ref{fig:SoftNLP}. 
By explicit computation, we find that the operators containing $J^{T2}_{\chi_\alpha \chi_\beta,\xi}$ do not mix into any of the local operators. The diagrams containing a single time-ordered product of $\mathcal{L}_{\xi}^{(1)}$ and any type of the local current vanish at the one-loop level for external states without soft fields and any number of collinear fields.  The reason is that the soft gluon field at $\order{\lambda}$ enters the Lagrangian only via the soft-field strength tensor with $\perp$ and $\nm$ components, $x^{\mu}_\perp \nnm{i}^\nu F_{\nu \mu_{\perp i}}$. Hence,  we observe that in the Feynman gauge, a diagram with single $\order{\lambda}$ Lagrangian insertion always contains the factor 
\be
  k_\alpha \left( g^{\alpha \nu}_{\perp i}\nnm{i}^\mu - \nnm{i}^\alpha g^{\mu \nu}_{\perp i} \right )(\nnm{j})_{\mu} \;,\nn
\ee
where $k$ denotes  the loop momentum and $(\nnm{j})_\mu$ comes from the soft vertex on the $j$-collinear line. No further $k$-dependent terms appear in the numerator because only the $\nm$ component of the soft momentum enters the collinear line and purely collinear interactions do not depend on the small component of the momentum.  
The one-loop soft loop integral depends on two vectors $\nnm{i}$ and $\nnm{j}$, so any tensor integral can be reduced to a combination of these vectors and a metric tensor. After the tensor reduction of the loop integral, the numerator terms with  $k \to \nnm{i}$  vanish by definition of the light-cone coordinates. If $k \to \nnm{j}$ then the total result is zero because of the anti-symmetric Feynman rule obtained from the soft gluon field-strength tensor. 

In summary, the time-ordered product operators with $\order{\lambda}$ Lagrangians do not mix into local currents. The renormalization factor of the mixing of the time-ordered products containing the $\order{\lambda}$ Lagrangian with themselves is given by the $Z$-factor of its local component. 
 
\section{Combined result}\label{sec:result}

In this section we discuss the combination of the collinear and soft contributions to the anomalous dimension.
As concluded above, at fermion-number two we can focus on current-current contributions.
We found that both the  collinear and soft contributions can be summarized in a universal way, given by
 Eq.\,\eqref{eq:Zcoll} and Eq.\,\eqref{eq:Zsoft}, respectively.
In particular, the total soft contribution, summed over all pairs of collinear directions $i, j$ with $i\not= j$, takes the form
\be
  \delta Z^s_{PQ}(x,y) = -\delta_{PQ}\delta(x-y) S
\ee
with 
\be
  S = \frac{\als}{4\pi}\sum_{i,j=1}^N(1-\delta_{ij})\sum_{l=1}^{n_i}\sum_{k=1}^{n_j}\frac{{\bf T}_{i_l}\cdot{\bf T}_{j_k}}{2}\left\{\frac{2}{\epsilon^2} +\frac{2}{\epsilon}\ln\left(\frac{-\mu^2 s_{ij}x_{i_l}x_{j_k}}{p_{i_l}^2p_{j_k}^2}\right)\right\}\,.
\ee
Notice that for identical building blocks, a symmetrization needs to be performed as discussed in the collinear case.
We can write the logarithm as a sum of three terms involving $-s_{ij}x_{i_l}x_{j_k}/\mu^2$, $\mu^2/(-p_{i_l}^2)$, and $\mu^2/(-p_{j_k}^2)$, respectively.
The last two terms are identical after renaming $i,l\leftrightarrow j,k$, thus we obtain
\be
   S = \frac{\als}{4\pi}\sum_{i,j}(1-\delta_{ij})\sum_{l,k}{\bf T}_{i_l}\cdot{\bf T}_{j_k}\left\{\frac{1}{\epsilon^2} +\frac{1}{\epsilon}\left[\ln\left(\frac{-s_{ij}x_{i_l}x_{j_k}}{\mu^2}\right)+2\ln\left(\frac{\mu^2 }{-p_{j_k}^2}\right)\right]\right\}\,.
\ee
Colour-neutrality of the entire $N$-jet operator implies 
$\sum_j\sum_k {\bf T}_{j_k}=0$. We can use this to rewrite $S$ as 
\be
   S = \frac{\als}{4\pi}\sum_{i,j}\sum_{l,k}{\bf T}_{i_l}\cdot{\bf T}_{j_k}\left\{\frac{1}{\epsilon}\ln\left(\frac{-s_{ij}x_{i_l}x_{j_k}}{\mu^2}\right)(1-\delta_{ij})-\delta_{ij}\left[\frac{1}{\epsilon^2}+\frac{2}{\epsilon}\ln\left(\frac{\mu^2 }{-p_{j_k}^2}\right)\right]\right\}\,.
\ee
When combining this with the collinear result in Eq.~\eqref{eq:Zcoll}, we 
find that the regulator-dependent terms cancel, as expected. This is a consequence of the colour conservation and our assumption that the operator is a colour singlet. The cancellation serves as a consistency check proving that the $N$-jet operator matrix elements have the correct IR behaviour and no further basis operators, in particular with soft building blocks, are necessary. 
Therefore, all current-current contributions to the $Z$-factor can be summarized as
\bea
  \delta Z_{PQ}(x,y) &=& \delta Z^s_{PQ}(x,y)+\delta Z^c_{PQ}(x,y)\nn\\
  &=& \delta_{PQ}\delta(x-y)\,\frac{\als}{4\pi}\sum_{i,j}\sum_{l,k}{\bf T}_{i_l}\cdot{\bf T}_{j_k}\Bigg\{\left[\frac{1}{\epsilon^2}+\frac{1}{\epsilon}\ln\left(\frac{\mu^2}{-s_{ij}x_{i_l}x_{j_k}}\right)\right](1-\delta_{ij}) \nn\\
  && -\delta_{ij}\delta_{lk}\frac{c_{i_l}}{\epsilon}\Bigg\} 
   + \sum_i\delta^{[i]}(x-y)\frac{\gamma^i_{PQ}(x,y)}{\epsilon}\;.
\eea
From this result we obtain the anomalous dimension matrix
\bea\label{eq:main}
\Gamma_{PQ}(x,y) &=& \delta_{PQ}\delta(x-y)\!\left[ 
-\gamma_{\textrm{cusp}}(\als)\sum_{i<j}
\sum_{l,k}{\bf T}_{i_l}\cdot{\bf T}_{j_k}\hspace{-0.02cm}
\ln\left(\frac{-s_{ij}x_{i_l}x_{j_k}}{\mu^2}\right) 
+ \sum_i\sum_l \gamma_{i_l}(\alpha_s) \right] 
\nn\\
  && + \,2\sum_i\delta^{[i]}(x-y)\gamma^i_{PQ}(x,y)\,,
\eea
where $\gamma_{\textrm{cusp}}(\als) = \frac{\als}{\pi}$, $\gamma_{i_l}(\alpha_s)\equiv -\frac{\als}{2\pi}{\bf T}_{i_l}^2 c_{i_l}=-\frac{3\als}{4\pi}C_F \, (0)$ for collinear quark (gluons), and the last line captures the off-diagonal contributions computed above.

This expression summarizes the main result of this work. We have checked that its form persists for all possible current-current contributions up to $\order{\lambda^2}$, beyond the $F=2$ operators considered here. Operator mixing and non-diagonal contributions with respect to collinear momentum fractions always enter via the collinear contributions $\gamma^i_{PQ}(x,y)$.

As a cross-check, Eq.~\eqref{eq:main} reduces to the leading-power result 
(\ref{eq:LPanomalousdim}) when there is only a single building
block in each collinear direction (i.e. $l,k= 1$, $x_{i_l},x_{j_k}\to 1$), 
such that in the notation used above $\delta(x-y)\equiv\prod_i\prod_{k>1}\delta(x_{i_k}-y_{i_k})\to 1$
is an empty product equal to unity. Furthermore, possibly non-diagonal contributions encapsulated in $\gamma^i_{PQ}$
vanish at leading power.\footnote{Note that we use a different normalization for the gluonic building block compared to Ref.~\cite{Becher:2014oda}, which affects
$\gamma_{i_l}(\alpha_s)$. At leading power, it is easy to see that the results agree when taking the different convention into account.}

In this work, we consider the case in which one of the collinear directions contains two fermionic building blocks (direction $i$, say).
At ${\cal O}(\lambda)$, there is only a single type of operators of this kind, given by the product of $J_i=J^{B1}_{\chi\chi}(t_{i_1},t_{i_2})$ defined in Eq.\,\eqref{eq:Jxixi}
for the direction labelled by $i$ and leading-power building blocks for all other $N-1$ directions $J_{j\not=i}=J_j^{A0}$.
In this case, the anomalous dimension is off-diagonal in the collinear momentum fractions in direction $i$,
\be
\sum_{j=1}^N\delta^{[j]}(x-y)\frac{\gamma^j_{PQ}(x,y)}{\epsilon} 
\to \frac{1}{\epsilon}\,\gamma^i_{\chi\chi,\chi\chi}(x_{i_1},y_{i_1})\,,
\ee
where the right-hand side is given by Eq.~\eqref{eq:gammaxixi}, and we have 
used $\gamma^j_{PQ}(x,y)= 0$ for all leading-power building
blocks $j\not= i$. Furthermore the product of delta functions for the $N-1$ other directions 
$\delta^{[i]}(x-y)\equiv\prod_{j\not=i}\prod_{k>1}\delta(x_{j_k}-y_{j_k})\to 1$
also collapses to unity.

At ${\cal O}(\lambda^2)$, there are two cases. Let us first consider the case 
that the direction $i$ which we choose to carry fermion-number two 
encompasses itself the ${\cal O}(\lambda^2)$ suppression, i.e. it is 
represented by one of the three operators in Eq.\,\eqref{eq:Jxixi2},
$J_i\in \{J^{B2}_{\chi\partial\chi},J^{B2}_{\partial(\chi\chi)},
J^{C2}_{{\cal A}\chi\chi}\}$.
Then the other $N-1$ directions have to contain leading-power building blocks, as before. The structure of the anomalous dimension follows
directly from Eq.\,\eqref{eq:Zmatrix}, and leads to operator mixing,
\be
  \sum_j\delta^{[j]}(x-y)\frac{\gamma^j_{PQ}(x,y)}{\epsilon} \to \frac{1}{\epsilon}
  \left(\begin{array}{ccc}
  \gamma^i_{\chi\partial\chi,\chi\partial\chi} & \gamma^i_{\chi\partial\chi,\partial(\chi\chi)} & \gamma^i_{\chi\partial\chi,{\cal A}\chi\chi} \\
  0 & \gamma^i_{\partial(\chi\chi),\partial(\chi\chi)} & 0 \\
  0 & 0 & \gamma^i_{{\cal A}\chi\chi,{\cal A}\chi\chi} 
  \end{array}\right)\,
\ee
where the non-zero contributions are given in Sec.\,\ref{sec:collLambda2} (specifically Eqs.\,\eqref{eq:gammaJxidxiJxidxi}, \eqref{eq:gammaJxidxiJAxixi} for the first
and Eq.\,\eqref{eq:gammaJAxixiJAxixi} for the last row, and $\gamma^i_{\partial^\mu(\chi\chi),\partial^\nu(\chi\chi)}=g_{\perp}^{\mu\nu}\gamma^i_{\chi\chi,\chi\chi}$ is
related to the ${\cal O}(\lambda)$ result Eq.\,\eqref{eq:gammaxixi}). The anomalous dimension is diagonal with respect to the other $N-1$ directions.

The second case that can occur at ${\cal O}(\lambda^2)$ is that direction $i$ 
with $F=2$ is described by the ${\cal O}(\lambda)$ contribution 
$J_i=J^{B1}_{\chi\chi}(t_{i_1},t_{i_2})$, and one of the other $N-1$ 
directions, say direction $i'$, contributes an additional ${\cal O}(\lambda)$ 
suppression. The remaining $N-2$ directions must then be represented by 
a leading-power building block. Since we do not require direction $i'$ to have
a definite fermion number, there are more possibilities, in particular $J_{i'}\in\{J^{A1}_{\partial\chi}, J^{A1}_{\partial{\cal A}}, J^{B1}_{{\cal A}\chi},
J^{B1}_{\cal AA}, J^{B1}_{\chi\chi}, J^{B1}_{\bar\chi \chi}\}$ (plus hermitian conjugated operators). In this case we need in addition the corresponding 
anomalous dimension matrices $\gamma^{i'}_{PQ}$ for these operators.
They will be given in future work.

In summary, we have taken the first step in a systematic investigation 
of the anomalous dimension of subleading power $N$-jet operators in view 
of resummation of logarithmically enhanced terms in partonic cross 
sections beyond the leading power. We provide an explicit result at the one-loop 
order for fermion-number two $N$-jet operators.  
In a forthcoming paper we will present results at ${\cal O}(\lambda)$, 
${\cal O}(\lambda^2)$ for general $N$-jet operators.

\subsubsection*{Acknowledgements} 
This work has been supported by the BMBF grant no. 05H15WOCAA.

\begin{appendix}

\section{Conventions}
\label{sec:conventions}

\bi
\item Collinear directions $\nnp{i}$, $i=1,\dots, N$ with $\nnm{i}\cdot\nnm{i}=\nnp{i}\cdot\nnp{i}=0$, $\nnm{i}\cdot\nnp{i}=2$.
Any momentum can be decomposed as
\be
  p^\mu = \frac12 \nnp{i}p \, \nnm{i}^\mu + \frac12 \nnm{i}p \, \nnp{i}^\mu + p_{\perp i}^\mu\,.
\ee
\item The different components of collinear momentum $p_i$ scale as  
$(\nnp{i} p_i,\nnm{i} p_i,p_{i\perp i}^\mu  )\sim 
(\lambda^0, \lambda^2,\lambda)$.
\item $n_i$ building blocks in direction $i$, labelled by $i_k$, 
$k=1,\dots,n_i$.
\item Abbreviation $s_{i j} =\frac{1}{2} (\nnm{i} \cdot\nnm{j}) P_i P_j$
\item Operators $J^{An}$, $J^{Bn}$, $J^{Cn}$ with one, two, three building blocks, respectively, and power suppression ${\cal O}(\lambda^n)$.
Here we count $J_\chi^{A0}=\chi_i=W_i^\dag\xi_i$ and 
$J_{\cal A}^{A0}={\cal A}^\mu_{\perp i} = W_i^\dag [iD_{\perp i}^\mu W_i]$ as leading power ($n=0$) for a collinear quark
and gluon, respectively. The power suppression of all other operators is then counted relative to the leading power.
\item Colour-space operator for parton labelled by $i_k$ is 
${\mathbf T}_{i_k}$ and colour conservation 
\be
\sum_{i=1}^{N}\sum_{k=1}^{n_i}{\mathbf T}_{i_k}=0\,.
\ee
\item We define $\als = \frac{g_s^2}{4\pi}$ and $\tilde{\mu}^2 = 
\mu^2 \,e^{\gamma_E}/(4\pi)$.
\item Covariant derivatives
\bea
  iD_{\perp i}^\mu&=&i\partial^\mu_\perp+g_sA_{\perp i}^\mu(x),\nn\\
  i\nnp{i}D_i&=&\nnp{i}(i\partial+g_sA_i(x))\,, \nn\\
  i\nnm{i}D_i&=&\nnm{i}(i\partial+g_sA_i(x)+g_sA_s(x_{i-}))\,, \nn\\
  iD_s&=&i\partial+g_sA_s(x)\quad \mbox{(on soft fields)}\,,\nn\\
  i\nnm{i}D_s&=&\nnm{i}(i\partial+g_sA_s(x_{i-}))\quad \mbox{(on collinear fields)}\,.
\eea
\ei

\section{Redundant operators}
\label{sec:redundant}

\subsection{\boldmath
Redundant collinear covariant derivative $i\nnm{i}D_i$}
\label{sec:redundant1}

In this Appendix, we show that the operator 
$\nnm{i}{\cal A}_{i} = W_i^\dag i\nnm{i} D_i W_i-i\nnm{i} D_{s}$, 
that could potentially contribute to the
basis of collinear building blocks at (relative) ${\cal O}(\lambda)$, can be 
expressed in terms of the operator basis
discussed in Sec.\,\ref{sec:basis}, and is therefore redundant (see also 
Ref.~\cite{Marcantonini:2008qn} for some closely related discussion).

The equation of motion for the collinear gauge field with respect to the  
$i$-th collinear direction derived from the 
leading-power collinear Lagrangian \cite{Beneke:2002ni} reads
\be\label{eq:YMeom}
[i D_{\nu i}, G_i^{\mu\nu}] = 
ig_st^a\bar\xi_i\left(\nnm{i}^\mu t^a+\gamma_{\perp i}^\mu t^a\frac{1}{i\nnp{i} D_i}i\slashed{D}_{\perp i}+i\slashed{D}_{\perp i}\frac{1}{i\nnp{i} D_i}\gamma_{\perp i}^\mu t^a+\dots\right)\frac{\slashed{n}_{i+}}{2}\xi_i\,,
\ee
where $ig_s G_i^{\mu\nu}=[i D^{\mu}_i,i D^{\nu}_i]$ and the ellipsis stand for 
contributions involving $\nnp{i}^\mu$,
that will drop out below. In the remainder of this Appendix we will 
consistently omit the index $i$ for the collinear direction $i$. 
The covariant derivative
\be
iD^\mu(x) \equiv i\partial^\mu + g_sA ^\mu(x)+ g_s\nm 
\Aus(x_-)\frac{\np^\mu}{2}\,,
\ee
includes the multipole-expanded soft field in the $\nm$ projection, 
$i \nm  D$. 
Contracting the equation of motion with ${\np}_\mu$ and multiplying with
collinear Wilson lines from both sides gives,
\be
W^\dag[i D_{\nu},[i\np D,i D^\nu]]W = -2g_s^2 W^\dag t^a W 
\,\bar\xi t^a\frac{\slashed{n}_+}{2}\xi\,.
\ee
Next we use $\sum_a t^a_{ij}t^a_{kl} = \frac12(\delta_{il}\delta_{jk}-\frac13\delta_{ij}\delta_{kl})$ to rewrite the colour
ordering on the right-hand side (colour indices made explicit)
\be
(W^\dag[i D_{\nu},[i\np D,i D^\nu]]W)_{ij} = 
-g_s^2  \left(\delta_{il}\delta_{jk}-\frac13\delta_{ij}\delta_{kl}\right)  
\overline{\chi}_k \frac{\slashed{n}_+}{2} \chi_l \,.
\ee
Writing the scalar product over $\nu$ on the left-hand side in terms of 
collinear basis vectors, and using $ W^\dag i\np D W = i\np\partial$ 
to simplify gives
\bea
(i\np\partial)^2(W^\dag [i\nm  D W])_{ij} &=& 
-2i\partial_{\perp\nu}(i\np\partial {\mathcal A}^\nu_\perp)_{ij}
- 2[{\mathcal A}^\nu_\perp,
(i\np\partial {\mathcal A}_{\perp \nu})]_{ij} 
\nn\\
  && {} - 2g_s^2\left(\delta_{il}\delta_{jk}-\frac13\delta_{ij}\delta_{kl}\right) \overline{\chi}_k \frac{\slashed{n}_+}{2} \chi_l\,.
\eea
Next, we apply the inverse derivative operator formally given by 
$1/(i\np\partial)^2$. Note that $(i\np\partial)^2(W^\dag [i\nm  D W])_{ij}$ 
transforms covariantly under the soft gauge symmetry, but 
$(W^\dag [i\nm  D W])_{ij}$ does not, since the derivative acts only 
inside the bracket. However, on the left-hand side
we can replace $W^\dag [i\nm  D W]\to W^\dag [i\nm  D W]-f(x_-)$ with an 
arbitrary function $f(x_-)$. This can also be seen as a freedom
to add an integration constant when applying the inverse derivative operator. 
It can be fixed by the requirement of soft gauge covariance, and 
choosing $f(x_-) = g_s\nm \Aus(x_-)$ yields
\bea\label{eq:Aminus}
(\nm {\cal A})_{ij} &=& 
-\frac{2}{i\np\partial}(i\partial_{\perp\nu}{\mathcal A}^\nu_\perp)_{ij}
- \frac{2}{(i\np\partial)^2}[{\mathcal A}^\nu_\perp,
(i\np\partial {\mathcal A}_{\perp \nu})]_{ij} 
\nn\\
  && {} - \frac{2g_s^2}{(i\np\partial)^2}\left(\delta_{il}\delta_{jk}-\frac13\delta_{ij}\delta_{kl}\right) \overline{\chi}_k \frac{\slashed{n}_+}{2} \chi_l\,,
\eea
i.e. we can express the operator on the left-hand side in terms of other collinear building blocks. The previous equation receives corrections from 
the power-suppressed interactions in the SCET Lagrangian, which can be 
worked out in a similar manner. Leading-power redundant operators can 
always be removed iteratively from these further terms.

One peculiar property of this relation is that the soft field appears explicitly only on the left-hand side. 
We checked that the relation is indeed 
fulfilled in the matrix element with one soft and one collinear gluon. On the 
left-hand side, a 1PI diagram exists, where the soft gluon is attached directly
to the operator. In addition, a 1PR diagram where the soft gluon is emitted 
from the collinear line contributes. On the right-hand side, only a 1PR 
diagram exists, that agrees with the sum of the 1PI and 1PR contribution 
from the left-hand side. We also checked explicitly that the identity 
holds in the matrix element with one and two collinear gluons with $\perp$
polarization. 

\subsection{\boldmath Redundant soft covariant derivative $i\nnm{i}D_s$}

We now show that the soft covariant derivative $i\nnm{i}D_s$ when operating 
on collinear fields can be removed using the collinear equations of motion.
As before, we omit the label for the collinear direction in this section for 
brevity. Using the equation of motion for the collinear quark field we find
\be
i\nm D_s \chi = -\left[\nm{\cal A}
+(i \slashed{\partial}_\perp+\slashed{\mathcal A}_\perp)
\frac{1}{i \np \partial}(i\slashed{\partial}_\perp+\slashed{\mathcal A}_\perp)\right]\chi,
\ee
which yields an expression in terms of the operator basis discussed in 
Sec.\,\ref{sec:basis} after using the relation \eqref{eq:Aminus} 
for $\nm{\cal A}$.
A computation similar to the one in Sec.\,\ref{sec:redundant1}, starting from the 
YM equation of motion \eqref{eq:YMeom} with open index $\mu$ projected in $\perp$ direction yields (with colour indices $ij$ made explicit)
\bea
\left([i\nm D_s, {\cal A}_\perp^\mu]\right)_{ij} &=&  
\frac12 i\partial^\mu_\perp(\nm{\cal A})_{ij} + \frac12\left([{\cal A}_\perp^\mu,\nm{\cal A}]\right)_{ij}
   +\frac{1}{2i\np\partial}\left([(i\np\partial{\cal A}_\perp^\mu),\nm{\cal A}]\right)_{ij} \nn\\
  && + \frac{1}{i\np\partial}\left(\big[i\partial_\perp^\nu+{\cal A}_\perp^\nu,[i\partial_\perp^\mu+{\cal A}_\perp^\mu,i\partial_{\perp\nu}+{\cal A}_{\perp\nu}]\big]\right)_{ij}\nn\\
  && + \frac{g_s^2}{2i\np\partial}\left(\delta_{il}\delta_{jk}-\frac13\delta_{ij}\delta_{kl}\right)
  \Bigg( \bar\chi_k\gamma_\perp^\mu\frac{1}{i\np\partial}\left(\slashed{\cal A}_\perp\right)_{ll'}\frac{\slashed{n}_+}{2}\chi_{l'} \nn\\
  && + \bar\chi_{k'}\left(\not\!\!{\cal A}_\perp\right)_{k'k}\frac{1}{i\np\partial}\gamma_\perp^\mu\frac{\slashed{n}_+}{2}\chi_l 
     + 2\bar\chi_k\frac{i\partial_\perp^\mu}{i\np\partial}\frac{\slashed{n}_+}{2}\chi_l\Bigg)\;.
\eea

\section{Auxiliary functions entering the anomalous dimension}
\label{sec:aux}

For the anomalous dimension $Z^{c,i}_{{\cal A}\chi\chi,{\cal A}\chi\chi}$ 
at ${\cal O}(\lambda^2)$ we need also the anomalous dimension 
$Z^{c,i}_{{\cal A}\chi,{\cal A}\chi}$ at ${\cal O}(\lambda)$ as an input.
It can be obtained by computing the one-loop matrix element 
$\langle g_{a}(q)\bar{q}(p)| J^{B1}_{{\cal A}^\mu\chi}(x) |0 \rangle$
and we find
\be
\delta Z^{c,i}_{{\cal A}^\mu\chi_\alpha,{\cal A}^\nu\chi_\beta}(x,y) = 
-g_\perp^{\mu\nu}\delta_{\alpha\beta}\delta(x-y)X_{i_1i_2} + 
\frac{1}{\epsilon}\,
\gamma^i_{{\cal A}^\mu\chi_\alpha,{\cal A}^\nu\chi_\beta}(x,y)\,,
\ee
with $X_{i_1i_2}$ given by Eq.\,\eqref{eq:Xlocal} and
\bea
\gamma^i_{{\cal A}^\mu\chi_\alpha,{\cal A}^\nu\chi_\beta}(x,y)  &=& 
{} \frac{\als {\bf T}_{i_1}\cdot{\bf T}_{i_2}}{2\pi}\,
\Bigg\{  g_{\perp }^{\mu\nu} \delta_{\alpha\beta}\Bigg( \theta(x-y)\left[\frac{1}{x-y}\right]_+ + \theta(y-x)\left[\frac{1}{y-x}\right]_+ 
\nn\\
  && \hspace*{-2cm}{} - \frac{\theta(x-y)}{\bar y}\left(1 +\frac{\bar x(\bar x+\bar y)}{2x}\right) 
        - \frac{\theta(y-x)}{2y}\left( \bar x+\bar y\right)\Bigg)
\nn\\
  && \hspace*{-2cm}{} + \frac14\,([\gamma_\perp^\mu,\gamma_\perp^\nu])_{\alpha\beta} (x+y)\bar x\left( \frac{\theta(x-y)}{\bar y x} + \frac{\theta(y-x)}{y\bar x}\right) \Bigg\}
\nn\\
  && \hspace*{-2cm}{} -  \frac{\als ({\bf C_F}+{\bf T}_{i_1}\cdot{\bf T}_{i_2})}{4\pi}  \,\Bigg\{g_\perp^{\mu\nu}\delta_{\alpha\beta}\Bigg( \frac{\theta(x-\bar y)\bar x}{yx}(\bar x+\bar y) + \frac{\theta(\bar y-x)}{\bar y}(\bar x-y)\Bigg)
\nn\\
  && \hspace*{-2cm}{} +\frac12([\gamma_\perp^\mu,\gamma_\perp^\nu])_{\alpha\beta}\Bigg( \frac{\theta(x-\bar y)\bar x}{yx}(\bar x- y-1) + \frac{\theta(\bar y-x)}{\bar y}(\bar x-y)\Bigg) \Bigg\}
\nn\\
  && \hspace*{-2cm}{} +\frac{\als {\bf C_F} }{4\pi} \,\bar x \,(\gamma_\perp^\mu\gamma_\perp^\nu)_{\alpha\beta}\,,
\label{eq:ZJ12J12}
\eea
where ${\bf C_F} \equiv \frac16(1-3(\T_{i_1}+{\bf D}_{i_1})\cdot \T_{i_2})$
and we introduced the additional colour operator 
${\bf D}^b|a\rangle = d^{abc}|c\rangle$ related to the symmetric $d^{abc}$
symbol defined via $\{t^a,t^b\} = \frac{1}{3}\delta^{ab}+d^{abc}t^c$.
We checked that our result agrees with Refs.~\cite{Hill:2004if, Beneke:2005gs} 
after subtracting the soft-loop contributions to the ${\cal O}(\lambda)$ 
heavy-to-light current from the anomalous dimension computed in these references.
By computing the matrix element 
$\langle \bar{q}(p)| J^{B1}_{{\cal A}^\mu\chi}(x) |0\rangle$ we furthermore 
find
\be
  \delta Z^{c,i}_{{\cal A}^\mu\chi_\alpha,\partial^\nu\chi_\beta}(x,y) = 0\,.
\ee

The functions entering $Z^{c,i}_{\chi\partial\chi,\chi\partial\chi}$ and $Z^{c,i}_{\chi\partial\chi,\partial(\chi\chi)}$ in 
Eq.~(\ref{eq:gammaJxidxiJxidxi}) are given by
\bea
  \lefteqn{ M_{\chi_\alpha\partial^\mu\chi_\beta, \chi_{\alpha'}\partial^\sigma\chi_{\beta'}}(x,y)}\nn\\
  &=& - \left(\theta(x-y)\frac{\bar x}{\bar y}+\theta(y-x)\frac{x}{y}\right)  x\bar x \nn\\
  && \times\Bigg[  -\left(\frac{\gamma_\perp^\sigma\gamma_\perp^\nu}{x}+\frac{\gamma_\perp^\nu\gamma_\perp^\sigma}{y}\right)_{\!\alpha\alpha'}\left(\frac{\gamma_\perp^\mu\gamma_{\perp\nu}}{\bar x}\right)_{\!\beta\beta'}
   - \left(\frac{\gamma_\perp^\mu\gamma_{\perp\nu}}{x}\right)_{\!\alpha\alpha'}\left(\frac{\gamma_\perp^\sigma\gamma_\perp^\nu}{\bar x}+\frac{\gamma_\perp^\nu\gamma_\perp^\sigma}{\bar y}\right)_{\!\beta\beta'} \nn\\
  && -2\delta_{\alpha\alpha'}\left(\frac{\gamma_\perp^\mu\gamma_\perp^\sigma}{\bar x\bar y}\right)_{\!\beta\beta'}
  -2\left(\frac{\gamma_\perp^\mu\gamma_\perp^\sigma}{ x y}\right)_{\!\alpha\alpha'}\delta_{\!\beta\beta'}
  -\frac{g_\perp^{\mu\sigma} }{x\bar x}\left(\gamma_\perp^\rho\gamma_\perp^\nu\right)_{\alpha\alpha'}\left(\gamma_{\perp\rho}\gamma_{\perp\nu}\right)_{\beta\beta'}
  \Bigg] \nn\\
  && {} +\frac12 \Bigg(\theta(y-x)\frac{x(x-2y)}{y^2} + \theta(x-y)\frac{\bar x(\bar x-2\bar y)}{\bar y^2}\Bigg) 
  \times \Bigg[ \left(\gamma_\perp^\sigma\gamma_\perp^\nu\right)_{\alpha\alpha'}\left(\gamma_{\perp}^{\mu}\gamma_{\perp\nu}\right)_{\beta\beta'} \nn\\
  && 
  + \left(\gamma_\perp^\mu\gamma_\perp^\nu\right)_{\alpha\alpha'}\left(\gamma_{\perp}^{\sigma}\gamma_{\perp\nu}\right)_{\beta\beta'}
  + g_\perp^{\mu\sigma}\left(\gamma_\perp^\rho\gamma_\perp^\nu\right)_{\alpha\alpha'}\left(\gamma_{\perp\rho}\gamma_{\perp\nu}\right)_{\beta\beta'} \Bigg]\,, 
  \nn\\
  \lefteqn{  M_{\chi_\alpha\partial^\mu\chi_\beta, \partial^\sigma(\chi_{\alpha'}\chi_{\beta'})}(x,y) }\nn\\
  &=& - \left(\theta(x-y)\frac{\bar x}{\bar y}+\theta(y-x)\frac{x}{y}\right) 
x\bar x\nn\\
  && \times \Bigg[  \left(\frac{\gamma_\perp^\sigma\gamma_\perp^\nu}{x}+\frac{\gamma_\perp^\nu\gamma_\perp^\sigma}{y}\right)_{\alpha\alpha'}\left(\frac{\gamma_\perp^\mu\gamma_{\perp\nu}}{\bar x}\right)_{\!\beta\beta'}
   +2\left(\frac{\gamma_\perp^\mu\gamma_\perp^\sigma}{ x y}\right)_{\!\alpha\alpha'}\delta_{\beta\beta'}
  \Bigg] \nn\\
  && {} +\frac12 \Bigg(\theta(y-x)\frac{x(\bar xy+y- x)}{y^2} + \theta(x-y)\frac{\bar x^2}{\bar y}\Bigg) 
  \times \Bigg[ \left(\gamma_\perp^\sigma\gamma_\perp^\nu\right)_{\alpha\alpha'}\left(\gamma_{\perp}^{\mu}\gamma_{\perp\nu}\right)_{\beta\beta'} \nn\\
  && 
  + \left(\gamma_\perp^\mu\gamma_\perp^\nu\right)_{\alpha\alpha'}\left(\gamma_{\perp}^{\sigma}\gamma_{\perp\nu}\right)_{\beta\beta'}
  + g_\perp^{\mu\sigma}\left(\gamma_\perp^\rho\gamma_\perp^\nu\right)_{\alpha\alpha'}\left(\gamma_{\perp\rho}\gamma_{\perp\nu}\right)_{\beta\beta'} \Bigg]\,.
\eea

The functions entering $Z^{c,i}_{\chi\partial\chi,{\cal A}\chi\chi}$ are 
given by
\bea
   K_{1,\alpha\alpha'\beta\beta'}^{ \mu\nu}(x,y_2,y_3)&\equiv& \delta_{\beta\beta'} G_{\alpha\alpha'}^{\mu\nu}(x,y_1,y_2)
  +\delta_{\alpha\alpha'}G_{\beta\beta'}^{\mu\nu}(\bar x,y_1,y_3) \nn\\
  && - H_{1, \alpha\alpha'\beta\beta'}^{\mu\nu}(x,y_1,y_2)-H^{\mu\nu}_{1,\beta\beta'\alpha\alpha'}(\bar x,y_1,y_3)-J^{\mu\nu}_{\alpha\alpha'\beta\beta'}(x,y_2,y_3)\,,
\nn\\[0.2cm]
  K_{2,\alpha\alpha'\beta\beta' }^{\mu\nu}(x,y_1,y_2) &\equiv& 2\delta_{\beta\beta'} F_{\alpha\alpha'}^{\mu\nu}(x,y_1,y_2) 
    - H^{\mu\nu}_{1,\alpha\alpha'\beta\beta'}(x,y_1,y_2) 
\nn\\
  && {} - H^{\mu\nu}_{2,\alpha\alpha'\beta\beta'}(x,y_1,y_2)
     + \frac12 I^{\mu\nu}_{\alpha\alpha'\beta\beta'}(x,y_1,y_2) \,,
\eea
where the contribution from diagram $(b, ii)_B$ and $(b, i)_B$ can be 
expressed in terms of
\bea
  G_{\alpha\alpha'}^{\mu\nu}(x,y_1,y_2)&\equiv&\frac{1}{1-y_3}\frac{1}{\bar x-y_3}\left(\theta(x-y_2)\theta(\bar x-y_3)\frac{\bar x-y_3}{y_1}+\theta(y_2-x)\frac{x}{y_2}\right)\nn\\
  && \times \left(-4xg_\perp^{\mu\nu}+(x-y_2+y_1)\gamma_\perp^\mu\gamma_\perp^\nu\right)_{\alpha\alpha'}\,.
\eea
The diagrams $(b, ii)_F$ and $(b, i)_F$ give
\bea
  F_{\alpha\alpha'}^{\mu\nu}(x,y_1,y_2)&\equiv&\frac{1}{1-y_3}\frac{1}{\bar x-y_3}\left(\theta(x-y_1)\theta(\bar x-y_3)\frac{\bar x-y_3}{y_2}+\theta(y_1-x)\frac{x}{y_1}\right)\nn\\
  && \times \left(2xg_\perp^{\mu\nu}- y_1\gamma_\perp^\mu\gamma_\perp^\nu\right)_{\alpha\alpha'}\,.
\eea
The diagrams $(c)_V$ and $(c)_V'$ give
\bea
  H^{\mu\nu}_{1,\alpha\alpha'\beta\beta'}(x,y_1,y_2) &\equiv& \left(\theta(x-y_1-y_2)\frac{\bar x}{y_3}+\theta(y_1+y_2-x)\frac{x}{y_1+y_2}\right)\nn\\
  && \times \left(\delta_{\beta\beta'}\frac{2\bar x}{y_1+y_2}(\gamma_\perp^\mu\gamma_\perp^\nu)_{\alpha\alpha'}+\frac{x}{y_1+y_2}(\gamma_\perp^\rho\gamma_\perp^\nu)_{\alpha\alpha'}(\gamma_\perp^\mu\gamma_{\perp\rho})_{\beta\beta'}\right)\nn\\
  H^{\mu\nu}_{2,\alpha\alpha'\beta\beta'}(x,y_1,y_2) &\equiv& \left(\theta(x-y_1-y_2)\frac{\bar x}{y_3}+\theta(y_1+y_2-x)\frac{x}{y_1+y_2}\right)\frac{x}{x-y_1}
\nn\\
  && \times \,(\gamma_\perp^\nu\gamma_\perp^\rho)_{\alpha\alpha'}(\gamma_\perp^\mu\gamma_{\perp\rho})_{\beta\beta'}\,.
\eea
The diagrams $(c)_F$ and $(c)_F'$ give
\bea
  I^{\mu\nu}_{\alpha\alpha'\beta\beta'}(x,y_1,y_2) &\equiv& 
  \Bigg(-\theta(x-y_1)\theta(\bar y_3-x)\frac{x^2\bar y_1+\bar x^2\bar y_3-\bar y_1\bar y_3}{\bar y_1 y_2\bar y_3} \nn\\
  && {} + \theta(y_1-x)\frac{x^2}{y_1\bar y_3} + \theta(x-\bar y_3)\frac{\bar x^2}{\bar y_1 y_3}\Bigg)\, 
  \Bigg\{ \frac{x+y_1}{x-y_1}(\gamma_\perp^\nu\gamma_\perp^\rho)_{\alpha\alpha'}(\gamma_\perp^\mu\gamma_{\perp\rho})_{\beta\beta'}\nn\\
  && {} + g_\perp^{\mu\nu} (\gamma_\perp^\sigma\gamma_\perp^\rho)_{\alpha\alpha'}(\gamma_{\perp\sigma}\gamma_{\perp\rho})_{\beta\beta'} 
  + (\gamma_\perp^\mu\gamma_\perp^\rho)_{\alpha\alpha'}(\gamma_\perp^\nu\gamma_{\perp\rho})_{\beta\beta'} \Bigg\}\;,
\eea
and the diagram $(c)_B$ yields
\bea
J^{\mu\nu}_{\alpha\alpha'\beta\beta'}(x,y_2,y_3) &\equiv& 
  \Bigg\{\frac12\,\Bigg(- \theta(x-y_2)\theta(\bar y_3-x)\frac{x^2\bar y_2+\bar x^2\bar y_3-\bar y_2\bar y_3}{\bar y_2 y_1\bar y_3} \nn\\
  && \hspace*{-2cm} {} + \theta(y_2-x)\frac{x^2}{y_2\bar y_3} + \theta(x-\bar y_3)\frac{\bar x^2}{\bar y_2 y_3}\Bigg) 
  \Bigg[ (\gamma_\perp^\nu\gamma_\perp^\rho)_{\alpha\alpha'}(\gamma_\perp^\mu\gamma_{\perp\rho})_{\beta\beta'}\nn\\
  && \hspace*{-2cm} {} + g_\perp^{\mu\nu} (\gamma_\perp^\sigma\gamma_\perp^\rho)_{\alpha\alpha'}(\gamma_{\perp\sigma}\gamma_{\perp\rho})_{\beta\beta'} 
  + (\gamma_\perp^\mu\gamma_\perp^\rho)_{\alpha\alpha'}(\gamma_\perp^\nu\gamma_{\perp\rho})_{\beta\beta'} \Bigg] \nn\\
  && \hspace*{-2cm} {} + \delta_{\alpha\alpha'}(\gamma_\perp^\mu\gamma_\perp^\nu)_{\beta\beta'}\Bigg(\theta(x-y_2)\theta(\bar y_3-x)\frac{\bar x-\bar y_2}{\bar y_2y_1}-\theta(x-\bar y_3)\frac{\bar x}{\bar y_2y_3}\Bigg) \nn\\
  && \hspace*{-2cm} {} + \delta_{\beta\beta'}(\gamma_\perp^\mu\gamma_\perp^\nu)_{\alpha\alpha'}\Bigg(\theta(x-y_2)\theta(\bar y_3-x)\frac{ x-\bar y_3}{\bar y_3y_1}-\theta(y_2-x)\frac{ x}{y_2\bar y_3}\Bigg) \Bigg\}\,.
\eea
For $0<y_i<1$ the functions $K_{1(2)}$ are regular for all $0<x<1$.\footnote{
There are terms contributing to $K_2$ that can potentially be singular for $x\to y_1$, in particular
\bea
  \frac12 I^{\mu\nu}_{\alpha\alpha'\beta\beta'}(x,y_1,y_2) &\to& \frac12\frac{x+y_1}{x-y_1}\Bigg(- \theta(x-y_1)\theta(\bar y_3-x)\frac{x^2\bar y_1+\bar x^2\bar y_3-\bar y_1\bar y_3}{\bar y_1 y_2\bar y_3} \nn\\
  && {} + \theta(y_1-x)\frac{x^2}{y_1\bar y_3} \Bigg) 
  (\gamma_\perp^\nu\gamma_\perp^\rho)_{\alpha\alpha'}(\gamma_\perp^\mu\gamma_{\perp\rho})_{\beta\beta'}\,, \nn\\
  - H^{\mu\nu}_{2,\alpha\alpha'\beta\beta'} &\to& -\frac{x}{x-y_1}\theta(\bar y_3-x)\frac{x}{\bar y_3}(\gamma_\perp^\nu\gamma_\perp^\rho)_{\alpha\alpha'}(\gamma_\perp^\mu\gamma_{\perp\rho})_{\beta\beta'}\,.
\eea
One can check that the sum of both terms is regular for $x\to y_1$. 
(One can use that in this limit $\theta(\bar y_3-x)\to 1$ due to
the assumption $y_2>0$. Then using $\frac{x^2\bar y_1+\bar x^2\bar y_3
-\bar y_1\bar y_3}{\bar y_1 y_2\bar y_3}\to -\frac{x}{\bar y_3}$ for 
$x\to y_1$, the two terms in the first and second line combine to cancel 
the pole in the third line.) Furthermore, there are additional occurrences 
of $1/(x-y_i)$, but one can check that the $\theta$-functions multiplying 
them exclude the pole for $0<y_i<1$.
}

\end{appendix}

\bibliography{paper}

\end{document}